# Investigation of the dynamics of inner gas during bursting of a bubble at the free surface

Digvijay Singh, Arup Kumar Das*

*Department of Mechanical and Industrial Engineering, IIT Roorkee, India-247667*

*Corresponding author's email: arupdas80@gmail.com

## Abstract

In the present study, simulations are directed to capture the dynamics of evacuating inner gas of a bubble bursting at the free surface, using Eulerian based volume of fluid (VOF) method. The rate by which surrounding air rushing inside the bubble cavity through the inner gas evacuation is estimated and compared by the collapsing bubble cavity during the sequential stages of the bubble bursting at the free surface. Further, the reachability of inner gas over the free surface is evaluated by establishing the comparison of the same through various horizontal planes, lying at different altitudes above the unperturbed surface. The evacuating inner gas accompanies vortex rings, which entrains the surrounding gas-phase. During the successive stages of air entrainment, spatiotemporal characteristics of the vortex ring are obtained. At low Bond numbers (< 1), after comparing the phase contours of evacuating inner gas from the bubble cavity, the consequences at the axial growth of gas jet and the radial expansion of the jet tip is discussed separately. Furthermore, under the respiration process, the axial growth of rising inner gas over the free surface and the radial expansion of vortex rings of a bubble bursting at the free surface is compared with the quiescent surrounding air. At last, the effects of various possible asymmetric perforation of the bubble cap keeping the same Bo are studied. The cause of bent gas jet, as a consequence of perforation of the bubble cap, asymmetrically, is explained by plotting the velocity vectors.

## 1. Introduction:

The spontaneous occurrence of bubble bursting phenomena at the free surface does a crucial role in various natural and industrial processes. Involving into mass, momentum and heat transfer at the gas-liquid interface [1-2], transport of microorganisms and surface-active material from the liquid phase [3-4], dying of cells during cultivation in the bioreactors [5-7] and effervescence in carbonated drinks [8] are a few examples of a decisive interaction often recognized at the gas-liquid interface. Due to a wider horizon of its applications, such a complex multiscale phenomenon has been studied for long. So far, many studies have been devoted to exploring the physics of a solitary bubble life cycle at the free surface [9-13]. Usually, a bubble acquires the non-spherical shape after reaching to the free surface, popularly known as static shape, after that, the bubble cap, segregating the bubble from the atmosphere, reduces its thickness through the route of film drainage and eventually disintegrates into the number of the tiny film drops, and further, the remaining open cavity deforms under the action of surface tension driven capillary ripples, the formation of a liquid column, also called Worthington jet, takes place at the base of the cavity after merging of the ripples inside the bubble and subsequently, jet disintegrates to release the jet drops, these are the primary stages noticed in a bubble bursting at the free surface. Kientzler et al. [9] had conducted a photographic investigation to capture the stages of the bubble bursting at the free surface and found that bubble diameter influences, directly to the time taken in the bubble bursting and the formation of a liquid jet. Later, during the sequential stages of a bubble bursting at the sea surface, MacIntyre [10] had discussed the occurrence of the boundary layer driven flow along with the bubble cavity interface and established the temporal evolution of cavity collapse.

Releasing film and jet drops are linked to the marine aerosol at the sea surface, the relative contribution of the individual into the formation of aerosol had been published in the experimental work of Cipriano and Blanchard [14]. They found that water-to-air salt flux is mainly due to jet drops only. A detailed and, to some extent contradicting viewpoint can also be seen in literature [2, 4, 15]. But the importance of jet drops in producing sea aerosol is well established, which has resulted in plenty of work in exploring the jet drop dynamics. The very first jet drop follows the oft-quoted 1-to-10 rule, later, Blanchard [14] found the upper limit to support the rule, quantitatively, the bubble diameter lies up to 500 μm at seawater temperature ~ 22-26°C. Remarks to the successive drops, the second jet drop is ~8% bigger in size than the top one, and the following drops are further smaller and bigger, both actually generate [14, 16]. Temperature is one of the key factors besides the bubble diameter to predict the size of the jet drop. At low temperatures, the same bubble originates small jet drops [14]. Occurrence of jet drops, the first and the second jet drop releases at nearest and farthest from the unperturbed free surface, respectively, and the successive jet drops are at the intermediate heights [14, 16]. The relevant velocity of the individual jet drop is reported in Spiel [16], stated that the maximum velocity is linked to the top jet drop. In contrast, in the following drops, the initial velocity has fallen gradually. A series of efforts by various research groups [17-24] can be noticed in this century, enlightening different fluidic physics of bursting bubbles. These studies highlight several important factors like the evaporation of aerosol [18], the minimum size of the top jet drop [20], no-jet drop condition [21], universal scaling law of top jet drop [22], universal cavity and jet profile [24]. Deike et al. [12] reported Worthington jet velocity is showing the function of focusing capillary ripples' wavelength. They found decreasing jet velocity with the Bond number (Bo) since the wavelength of the focusing capillary ripples moves up with "Bo". Berny et al. [25] have established the number of releasing jet drops during bubble bursting for a wider range of "Bo" and Laplace number (La) and noticed no jet drop for $La < 500$. One can note that the list is not exhaustive for the research of bubble bursting at the free surface, and many more significant contributions are there in literature.

At the same time, some efforts have also been given to explore the understanding of bubble bursting in a cluster of bubbles [26-28]. Experimental work on flower-shaped bubbles around the bursting bubble is reported by Liger-Belair et al. [26]. Later on, a comprehensive and detailed discussion on the bubble bursting in a cluster of bubbles is reported in the numerical effort of Singh and Das [28], by solving the two-phase Navier Stokes equations using open source solver, Gerris [29]. They have found that, in the symmetric placement of bubbles around the collapsing one, the time induced in the development of Worthington jet after merging of the capillary ripples inside the bubble and the pinching mechanism to release the jet drops is considerably low relative to a solitary bubble bursting at the free surface. In the asymmetric placement of bubbles around, rising liquid jet bends towards the absence site of bubbles [26, 28].

Apart from such an exhaustive study, still, a lot of uncovered physics of bursting bubbles needs to be understood. For example, the inner gas of the bursting bubble plays a vital role in enhancing the taste of carbonated drinks. But no systematic effort in the list of the above literature has targeted the study of inner gas. The only mention of inner gas dynamics in literature can be found in the experimental demonstration of Al Dasouqi and Murphy [30-31], who have recorded the dynamics during the bursting of a smoke-filled bubble. Their observation highlights a burst of the bubble at the free surface, which generates gas jet as an outcome of the phenomenon, accompanied by multiple vortex rings. However, detailed documentation of inner gas dynamics and in-depth understanding of influencing parameters is not available till date. On the other hand, natures of inner gas spread under the influence of respiration and asymmetric pinch off are still open questions. Therefore, the present numerical study is intended to establish a comprehensive analysis of each stage of emerging inner gas of bubble bursting at the free surface. We show the diffusion of inner gas in the air after the bursting of the bubble at the free surface. Entrainment of air inside the bubble cavity is also tracked for a wide range of operating parameters. To understand the effect of breath in and out on the gas diffusion, simulations are performed using inflow and outflow at the rate of respiration air velocity. Towards the end, using three-dimensional simulations, efforts have been made to understand the consequences of asymmetric pinch-off of the bubble. This paper is organized in the following form; 1) introducing the phenomena and earlier studies devoted to exploring the hidden physics of bubble bursting at the free surface, 2) numerical formulation, discretization schemes, and grid sensitivity of the results, 3) direct and derived observations of simulation results, including critical comments and in-depth discussions, and 4) concluding remarks to summarize the gist of present work.

## 2. Numerical Methodology:

To perform the simulations, an open-source computational tool, Gerris [29] is used to solve first, the axisymmetric, and then full three dimensional, two-phase, incompressible Navier Stokes equations. The computational domain is discretized in a quadtree cartesian mesh system. To predict the interface, Eulerian based volume of fluid (VOF) scheme is employed. Grid adaptation based on the gradient of phase fraction (T) is considered to refine the interface locally. Earlier, Gerris has been used extensively to predict the complex interfacial phenomenon [12, 18, 32-35].

### Domain and boundary condition

Present work is related to the dynamics of the inner gas of the bursting bubble and its consequences at the free surface. The primary objective is to establish the fluidic stages of inner gas during the bursting of a gas bubble. The static shape of the bubble at the free surface is obtained by balancing the buoyancy and the capillary forces. The buoyancy force maintains the bubble emerging at the free surface while the capillary force keeps the bubble partially submerged inside the liquid pool. To obtain the static shape below the free surface, mathematically, the inertia, the gravity, and the capillary forces in the cylindrical coordinate system are balanced and nondimensionalized. Accordingly, for the portion of the bubble cavity immersed in the liquid phase, we get:

$$\frac{1}{r}\,\frac{d(r sin\varphi)}{dr} = 2 + Bo \times z \tag{1}$$

Where $\varphi$ denotes the angle subtended by the lines joining the interface normal and vertical centroidal axis, and Bo is Bond number.

On the other hand, force balance in the azimuthal direction across the bubble periphery gives:

$$dz/d\varphi \ = -\,dr/d\varphi \times tan\varphi \tag{2}$$

Simultaneous solution of Equation (1) and (2) gives the static shape of the bubble cavity below the free surface. The potion of the bubble static shape over the free surface is characterized by "h", which can be evaluated at any free surface by matching the tangential slope as:

$$dh/dr = \ -\tan(\phi_c) \tag{3}$$

where $\phi_c$ is denoting the azimuthal angle at which the interface of the bubble asymptotically matches the free surface.

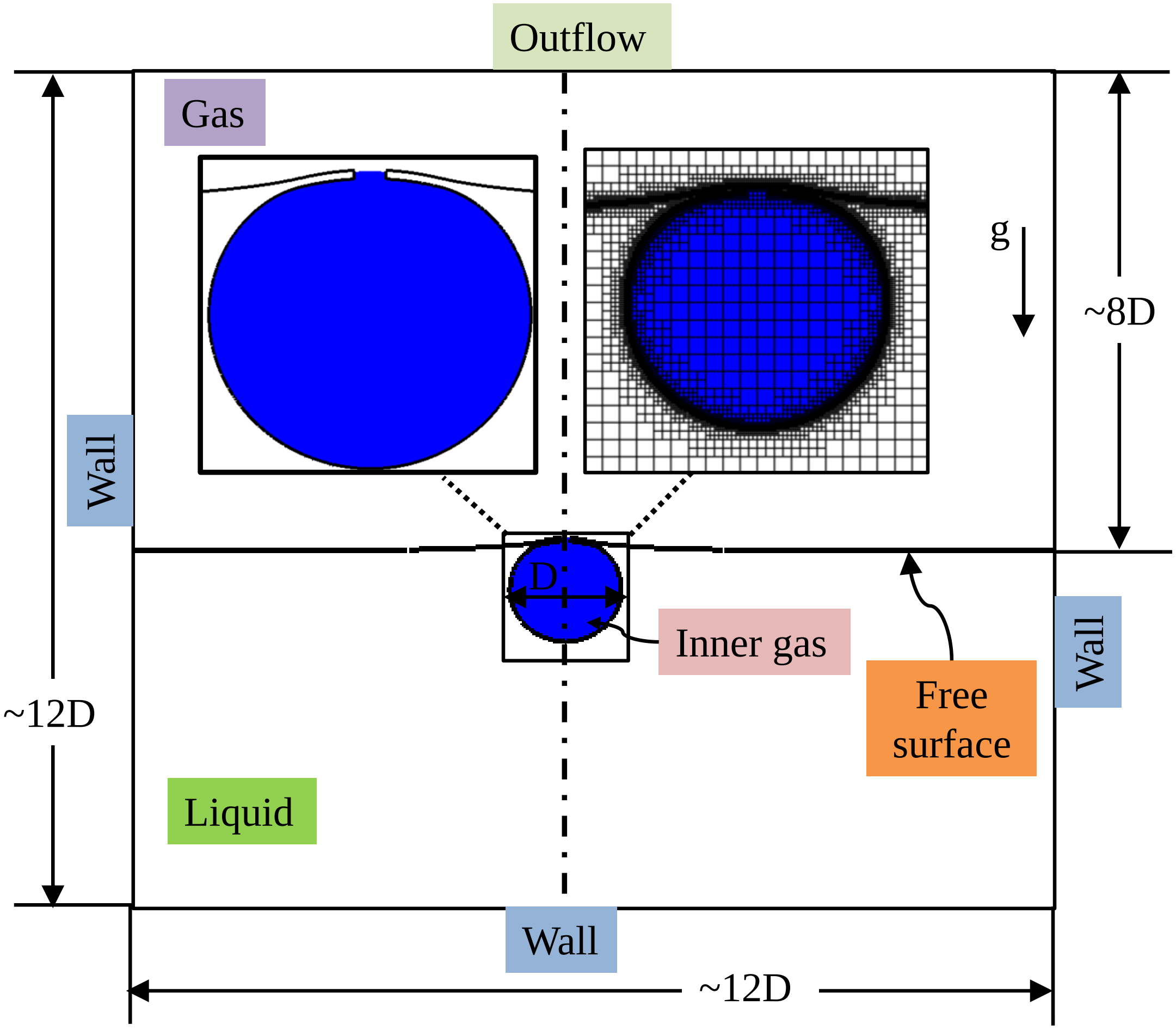

**Figure 1:** Schematic representation of computational domain and associated boundary condition to set up the simulations; boundary outflow to the top surface and remaining boundary faces are no-slip, no-penetration type, inset, an enlarged view of pierced static bubble and adaptive quadtree grid discretization.

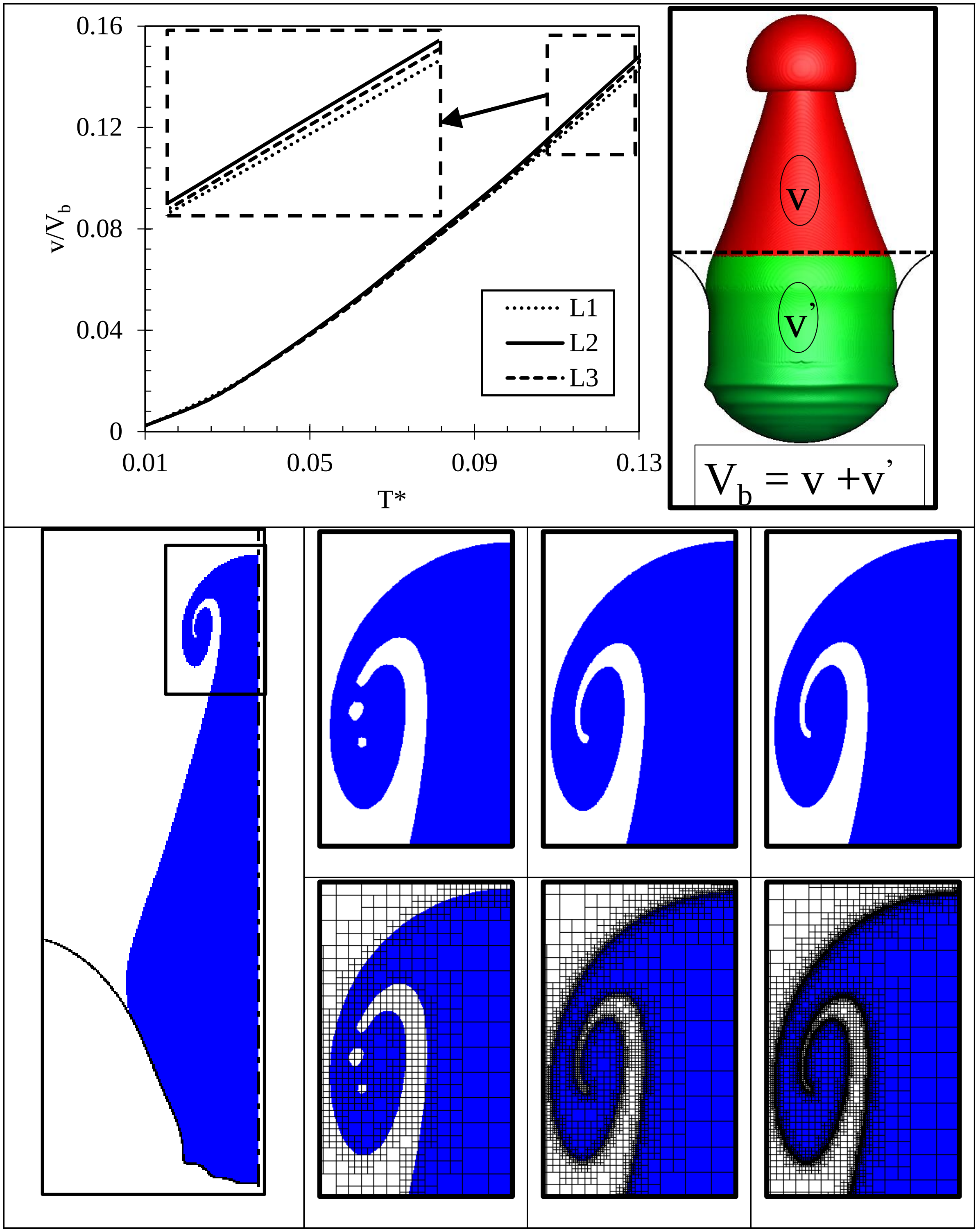

**Figure 2:** Grid independent study performed at several levels of refinement for axisymmetric simulations

We followed the same algorithm as mentioned above and elaborated in Singh and Das [28], to create a static shaped bubble at the free surface. Then we have set cap film thickness as ∼D/36 and considered axisymmetric puncture diameter as ∼D/20. Note that we do not attempt to resolve the break-up of the thin bubble cap numerically but start the simulation directly with the retracting opened cavity. A better resolution of the film at the nanometer level would have been quite an accurate prediction. But as our present focus is on the inner gas and towards the affordable computational cost, all the simulations have been started with ~D/36 thickness of the initial thin film. The computational domain associated with boundary conditions to set up the numerical simulations is illustrated in Figure 1. The discretized domain is locally refined around the interface with respect of the gradient of phase fraction, as shown in the inset of Figure 1.

To check the appropriateness of reported results, we conducted axisymmetric simulations at several grid sizes and reported the temporal evolution of inner gas volume above the free surface (v), nondimensionalized with the overall volume of the inner gas ($V_b$; shown in Figure 2). Here, time ($T^*$) has been nondimensionalized with capillary time scale ($\sqrt{\rho_l D^3/\sigma}$). One can note that for expelled inner gas volume above the free surface, all three grids (L1, L2, and L3; mentioned in Table 1) perform satisfactorily. Upon investigation on the smaller scale features for different grid sizes, we have observed that inner gas vortex shape (shown in Figure 2) saturates at L2 mesh size. Simulation run time and prediction improvements for different meshes are depicted in Table 1. Based on these observations, for axisymmetric simulations, grid discretization, L2, is considered a suitable grid size to conduct the simulations. In this grid configuration (level 13 in Gerris [36]), ~491 cells house in the diametrical plane of the initial bubble configuration. To improve the simulation time, during the run of the simulation, grids are adapted based on criteria of tracer gradient across the gas-liquid (T) and inner gas-rest phases ($T_i$) with a maximum level of 13. We have followed the same algorithm, mentioned in Singh and Das [28], which has already been validated with existing data.

**Table 1.** Numerical outcomes by performing the grid-independent study of axisymmetric simulations

| **Level** | **L1** | **L2** | **L3** |
|---|---|---|---|
| **Cell size, µm** | 14.64 | 3.66 | 1.83 |
| **Time, hr** | 0.42 | 6.13 | 16.93 |
| **Average % change in accuracy** | **L1→ L3:** 2.51 | **L2→ L3:** 1.41 | |

The governing equations to describe the physical system are incompressible mass conservation and two-phase momentum conservation, given below by equations (4) and (5), respectively. Similar sets of equations are already implemented in the framework of Gerris [36], an open-source solver, and the same has been used in the present context. The functional form of the equations are as follows:

$$\nabla.\boldsymbol{V} = 0 \tag{4}$$

$$\rho\left[\frac{\partial \boldsymbol{V}}{\partial t} + \boldsymbol{V}.\nabla\boldsymbol{V}\right] = -\nabla \mathrm{p} + \nabla.(2\mu\mathbf{D}) + \sigma \mathrm{k} \delta_s \mathbf{n} + \mathbf{f_b} \tag{5}$$

Here, $\boldsymbol{V} = u\hat{\imath} + v\hat{\jmath} + w\hat{k}$ is the fluid velocity, and $\rho$ & $\mu$ are denoting to density and viscosity, respectively. $\mathbf{D}$ is the deformation tensor, defined by $\frac{1}{2}\left(\frac{\partial u_i}{\partial x_j} + \frac{\partial u_j}{\partial x_i}\right)$, while $\sigma$, k, $\delta_s$ and $\mathbf{n}$ are indicating to surface tension coefficient, curvature, Dirac delta function, and unit normal vector, respectively. $\mathbf{f_b}$ is the body force due to gravity. Volume-fraction weighted average equations are used to calculate the fluid properties (density ($\rho$), viscosity ($\mu$)) in the two-phase system [29]:

$$\rho(\mathrm{T}) = \rho_1 \mathrm{T} + \rho_2(1 - \mathrm{T}) \tag{6}$$

$$\mu(\mathrm{T}) = \mu_1 \mathrm{T} + \mu_2(1 - \mathrm{T}) \tag{7}$$

Spatiotemporal instances of the interface are traced by the constitutive equation of the phase fraction (T):

$$\frac{\partial T}{\partial t} + \boldsymbol{V}.\nabla \mathrm{T} = 0 \tag{8}$$

Here, T is denoting the phase fraction, and the value lies in [0, 1]. A separate variable tracer, $T_i$ is used to define the inner gas of bubble and $T_i \in [0, 1]$. It is worth noting that the value of $T_i$ is non-zero only for the cells defining the bubble. The interface defined by $T_i$ is reconstructed by piece-wise linear interface construction (PLIC) scheme in the VOF system, the same as in T. To model the diffusion between inner gas ($T_i = 1$, $T = 0$) and air (($T_i = 0$, $T = 0$), height function-based interface reconstruction has been performed considering $T_i \times T$ as a variable. Here, entrainment of air in inner gas happens due to differential velocity based stress tensor with the same viscosity coefficient ($\mu_{air} = \mu_{inner\ gas}$).

The numerical schemes and corresponding discretized governing equations are given in the following discussion. The classical time-splitting projection method is used to discretize the momentum equation (5), and after decoupling the pressure term, we get the modified version of the momentum equation (9).

$$\rho_{n+\frac{1}{2}}\left[\frac{\mathbf{V}_* - \mathbf{V}_n}{t_* - t_n} + \mathbf{V}_{n+1}.\nabla\mathbf{V}_{n+1}\right] = \nabla.\left[\mu_{n+\frac{1}{2}}(\mathbf{D_n} + \mathbf{D}_*)\right] + (\sigma k \delta_s \boldsymbol{n})_{n+\frac{1}{2}} + \mathbf{f_b} \tag{9}$$

In this Equation (9), velocity advection term ($\mathbf{V}_{n+1}.\nabla\mathbf{V}_{n+1}$) is solved by the second-order un-split upwind scheme, as mentioned in Bell-Colella-Glaz [37], and Crank-Nicholson scheme (second-order accurate) [29] to viscous term (**D**). The constitutive equation of phase fraction (equation 8) is solved using the central difference scheme about $t_n$:

$$\left[\frac{\mathrm{T}_{n+\frac{1}{2}} - \mathrm{T}_{n-\frac{1}{2}}}{t_{n+\frac{1}{2}} - t_{n-\frac{1}{2}}}\right] + \nabla.(\mathrm{T_n}\mathbf{V}_\mathrm{n}) = 0 \tag{10}$$

The velocity correction ($\mathbf{V}_*$), to decouple the pressure term from the momentum equation, is defined as:

$$\mathbf{V}_{n+1} = \mathbf{V}_* - \frac{t_{n+1} - t_*}{\rho_{n+\frac{1}{2}}} \nabla p_{n+\frac{1}{2}} \tag{11}$$

Now, after taking the divergence of Equation (11), we get the pressure-Poisson's equation:

$$\nabla.\frac{t_{n+1} - t_*}{\rho_{n+\frac{1}{2}}} \nabla p_{n+\frac{1}{2}} = \nabla.\mathbf{V}_* \tag{12}$$

The numerical procedure to get the updated flow parameters is already discussed in Singh and Das [28]. Additional terms in the momentum equation, the surface tension is containing the curvature (k), calculated by height function model [38], and surface tension coefficient ($\sigma$), by the balanced force method [29].

## 3. Results and discussion:

Two-dimensional axisymmetric simulations with quadtree meshes have been performed to explore the dynamics of the inner gas after bursting of a static shaped bubble at the free surface [28]. Eulerian, the volume of fluid (VOF) based framework, as described in the previous section, has been adapted for simulations. A representative case, having Bond no., Bo = $\Delta\rho g D^2/4\sigma = 0.1725$, Laplace no., La = $\sigma\rho_l D/2\mu_l^2 = 4.14\times10^4$ and Morton no., Mo = $g\mu_l^4\Delta\rho/\rho_l\sigma^3 = 1\times10^{-7}$ is considered for the description of the dynamics of inner gas inside the bubble. Here, D represents the equivalent bubble diameter, $\rho_l$, and $\mu_l$ are denoting density and viscosity of the liquid phase. As described in several works of literature [1-6, 11, 18-21], present simulation also shows complex fluidic phenomena and interfacial evolution during the bursting of a bubble. It comprises the following stages; i) the rupture of thin liquid film separating the bubble from the atmosphere, ii) retraction of the orifice, created after rupture of the liquid film, iii) collapse of the bubble cavity during traversal of low-pressure capillary ripples, iv) formation of a high-speed liquid jet, also called Worthington jet, as a consequence of merging of the capillary waves at the base in bubble cavity, and subsequently v) pinching off of jet drops. It needs to be noted that besides interfacial evolution, the dynamics of inner gas to the bubble is worthy of study. Notably, in the case of carbonated drinks, taker experiences a distinctive, typically a pleasant fragrance around the nasal cavity before ingesting the drinks, which is the aroma of inner gas. Sometimes, such fragrances become irritant under the excessive presence of dissolved chemical species in the drinks. Recent studies by Al Dasouqi and Murphy [30-31] made efforts for understanding the dynamics of smoke-filled bubbles through experiments and observed the rising smoke jet associated with vortex rings. To further enrich the knowledge, the present study targets specifically the observation of inner gas dynamics from the above-mentioned numerical simulation. In the present study, capillary time, $\sqrt{\rho_l D^3/\sigma}$ is used to get non-dimensionalized time (T*). Here, $\rho_l$ , D and σ are denoting the density of the liquid, equivalent bubble radius, and coefficient of surface tension, respectively.

### 3.1: Sequences of inner gas release and characteristic features

The process of impulsive flow of the inner gas from the bubble cavity during bursting at the free surface is reported in Figure 3. An inner gas evolves in space and time during the sequential stages of the bursting bubble at the free surface. Initially, during the liquid film retraction, an inner gas produces a strong vertical upward-moving jet through the opening. During this process, the inception of a vortex ring is noticed, as shown in Figure 3 (c). It continuously entraps the surrounding air phase and grows with time, as reported in subsequent temporal inner gas contours in Figure 3. When the capillary waves reach the bottom of the bubble cavity (see Figure 3(e)), a tiny gas bubble pinches off into the liquid pool, after the confluence of waves, as shown in Figure 3 (f). An enlarged view of the entrapped tiny gas bubble along with grid discretization is showed beside. As a consequence of the confluence of capillary waves, a high-speed Worthington jet is produced, which pushes the inner gas from the base of the bubble cavity, as revealed in Figure 3 (h). After gaining a certain height, the liquid jet pinches off the drop, which travels faster than liquid and an inner gas jet. On reaching the tip of the gas jet, drop drags inner gas, and a beak-like inner gas bulge is produced (Figure 3 (m)). This thin vertical column of the inner gas follows the drop and extend its reach than the nominal vortex driven an inner gas phase. The axial growth of the gas jet over the free surface is displayed in Figure 4 to understand the reach of inner gas. Indeed, three distinct zones are recognized in which at the commencement, the inner gas evacuates parabolically until the top jet drop arrives at the gas jet tip. Subsequently, expedited linear vertical growth of jet tip is noticed due to drag behind the top jet drop, which achieves higher heights keeping disconnected inner gas vicinity around. Eventually, the droplet surrounded by the inner gas is detached, reflecting a sudden fall in the tip height. Following the same trend, again and again, expedited but lesser strength linear growth of inner gas tip happens in subsequent jet drops. In figure 4, the same has been exposed for the second jet drop. The same explanation can also be obtained from the velocity of the gas jet tip, shown in the inset of figure 4. Spiel [16] has already stated that the velocity of subsequent jet drop falls relatively to the previous droplet. It is to be noted that similar inner gas jet motion has been only qualitatively reported by Al Dasouqi and Murphy [30-31].

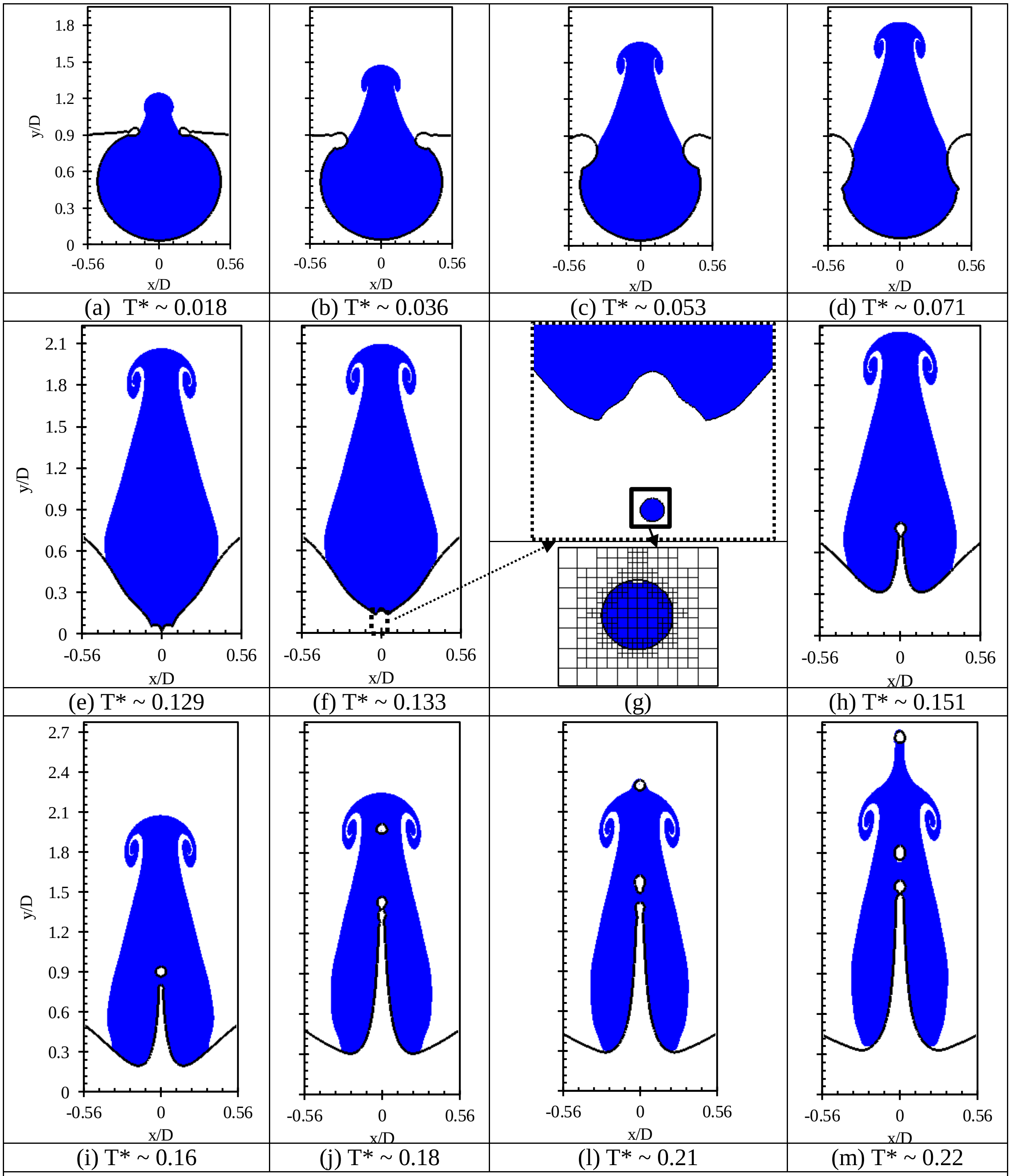

Figure 3: Temporal evolution of the inner gas of the bubble during the sequential stages of bursting at the free surface. Equivalent diameter, D = 1.8 mm, Bond number, Bo = 0.1725, Laplace Number, La = 4.14×104 and Morton Number, Mo = 1×10-7.

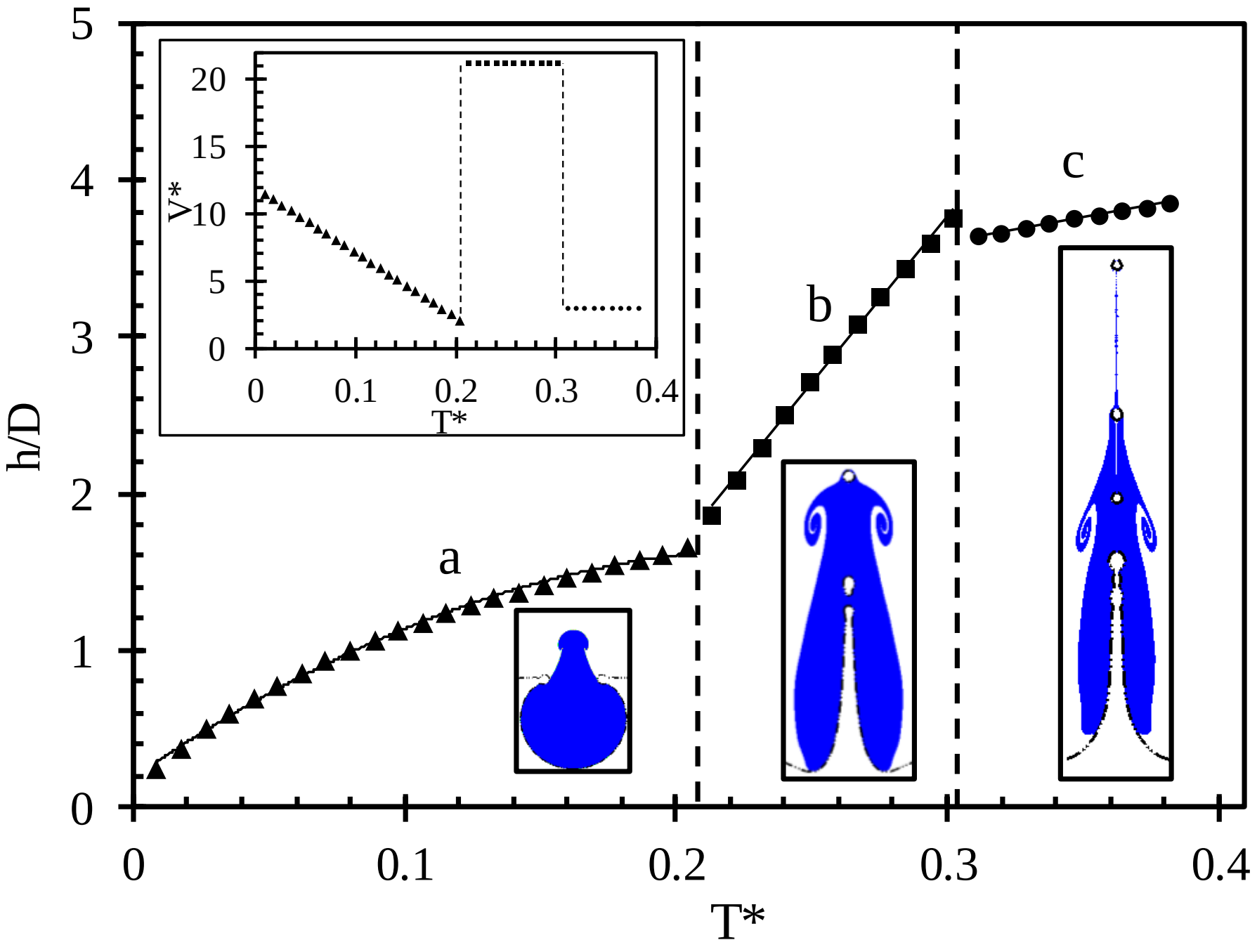

**Figure 4:** Temporal growth of the inner gas jet tip over the unperturbed free surface during the sequential stages of the bursting bubble, and the corresponding velocity is given in the inset.

### 3.2: Expulsion of inner gas and entrainment of air

At the base of the liquid jet, the collapsing cavity pumps out the gas inside the bubble and simultaneously, entrains some air below the unperturbed free surface. An effort has been made to keep the account of volumes inside the bubble cavity below the unperturbed free surface. As the volume of the bubble cavity will be occupied by inner gas and entraining air, a separate track has also been prepared for these volumes in the same plot. One can see a fast evacuation of inner gas due to bubble curvature related to high pressure than the atmosphere. Around T* ~ 0.3, complete evacuation of inner gas from the bubble cavity below the unperturbed free surface can be noticed. At the same time, atmospheric air starts entraining inside the bubble cavity, initially at a slower rate, which then picks up due to free space created during inner gas evacuation. Soon, the whole cavity is found to be occupied by entraining air only, which, subsequently, is also observed to be expelled out during neutralization of the cavity at the free surface. By summing up the account of inner gas and entrained air, the progression of bubble cavity with time can be obtained, which was initially found to be decreasing due to dominant and bubble curvature driven inertial evacuation of inner gas till T* ~ 0.16. After this transition period, entrained air diffusion takes the lead role in deciding the cavity volume, which increases with the increase of radial stretch of the bubble at the free surface. During this period, cavity volume increases, after T* ~ 0.3, the whole cavity is found to be occupied by entrained air. Bubble cavity nullifies into the unperturbed free surface by creating some dying waves. At this phase, wavelength and wave amplitude will decide the entrained air volume below the unperturbed free surface.

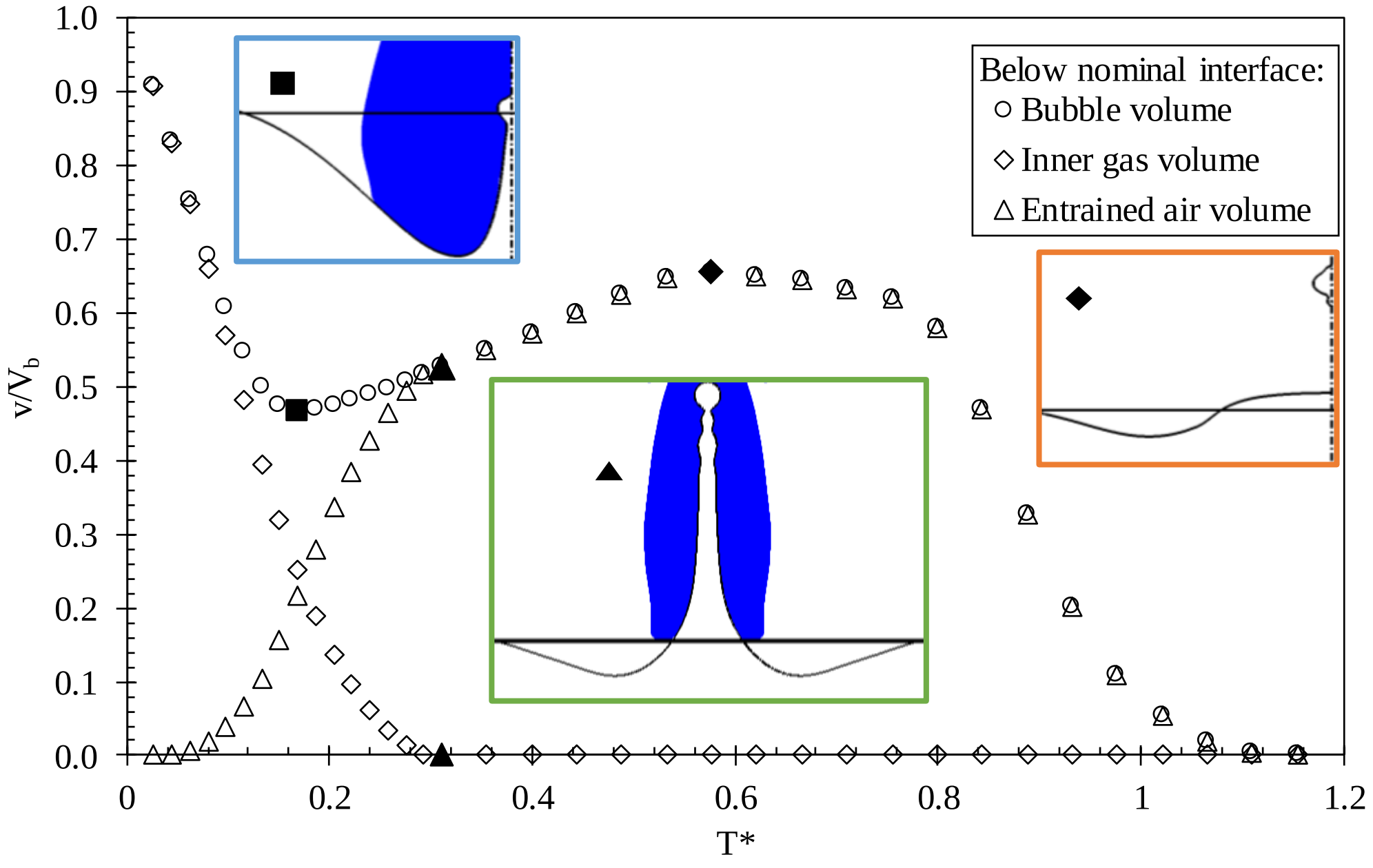

(a)

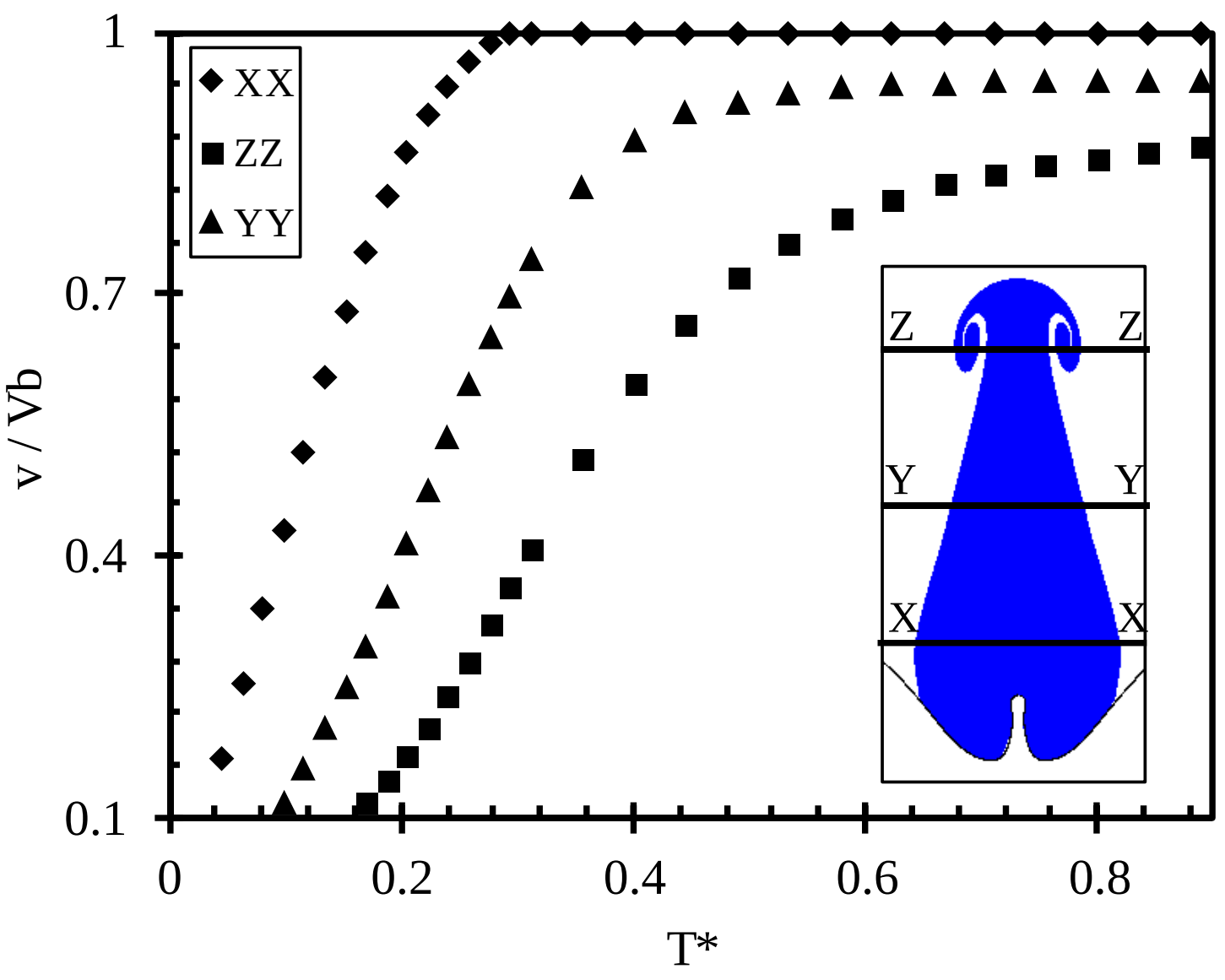

**(b)**

**Figure 5:** (a) The account of inner gas volume inside bubble cavity below the free surface and rushing of surrounding air into the bubble until the complete restoration of the free surface, (b) Comparison of inner gas crossing the plane XX, YY & ZZ.

Eventually, it will be leading asymptotically to zero entrained volume after reaching the peak around T* ~ 0.5. Hence, the volume account of bubble cavity below the free surface with time initially reduces due to high outward inertia of inner gas, then increases as a result of diffusion-driven entrainment and subsequently nullifies to zero after obtaining a peak value at which the cavity is filled with entrained air only. Phase contours of these volumes are also shown at three different locations in figure 5, to clearly explain the concept of change in the regime of cavity collapse.

Another critical parameter for the study can be the amount of inner gas crossing a particular plane parallel to the free surface. The volume of an inner gas crossing a plane at some height will dictate the reach of aroma or dissolved gas in the carbonated fizzy drinks like champagne after bubble bursting. To understand the same, effort has been made to estimate the reachability of inner gas at different heights of the plane above the free surface. XX plane is considered exactly at the unperturbed free surface. Subsequently, above ~0.5D and ~D from the unperturbed free surface, YY and ZZ planes are also chosen for the study. Figure 5 (b) is comparing the reach of inner gas over and above the planes with time. One can observe a swift expulsion of an inner gas over the planes due to the pumping action of the collapsing bubble. The inner gas first reaches the XX plane and then the YY plane. Finally, after elapsing a certain period, it reaches a ZZ plane. The slope in a variation of $v/V_b$ with T* (Figure 5 (b)) shows that the rate of expulsion of inner gas reduces upon moving up from the unperturbed free surface. It indicates that the effect of pumping action is more in the vicinity of the collapsing bubble, and the effectiveness of aroma reduces as one moves away from the free surface. In the later part, before the complete evacuation of inner gas, above a surface, rate of expulsion asymptotes with time due to lateral diffusion. In the plane, XX, the effect of diffusion is less, which can be observed from a fast change in the nature of expulsion from linearly increasing to complete saturated expulsion ($v/V_b = 1$). In YY and ZZ planes, due to higher diffusion in the lateral direction, inner gas rise above the designated plane slows down as compared to expulsion at the lower plane. Gradually, in upper planes, inner gas expulsion transforms to apparent asymptotic slow expulsion nature, which over a long period (not shown in Figure 5) leads towards complete expulsion. One can also note that the transition from fast evacuation to slow expulsion regime happens at a lower value of $v/V_b$ for higher horizontal planes. This can be explained from an increase of gas diffusion tendency in the lateral direction as the height from free surface increases.

### 3.2: Entrainment of air in the inner gas vortex

The impulsive flow of inner gas entrains the irrotational atmospheric air by forming a vortex ring in the vicinity of the inner gas jet tip. The spatiotemporal instances of the tip of the vortical flow of entraining air are shown in Figure 6. The process of air entrainment inside the vortex ring can be seamlessly connected through different stages, which are mentioned below.

**Stage a:** In the beginning, the vortex ring grows in size without entraining air, helping the inner gas to spread in the outward direction. In this process, around T* ~ 0.02, atmospheric air is dragged in by vortex motion from its bottom at the inner side. It initiates the air entrainment in the vortex ring. Initially, for the entrainment, no radial growth (*r*) is observed until T* ~ 0.03. It can be visualized from Figure 6 at the beginning with a constant value of *r* over time and decreasing nature with time for the gap between the entrainment tip and inner gas jet tip ($h_{vt}$). At this stage, the tip of entrained air shows a fast rise in the axial distance as compared to the unperturbed pool height (shown in inset). It needs to be noted that due to the parabolic growth of the inner gas jet tip with time (figure 4), a steady upward movement of entrained air tip will be sustained.

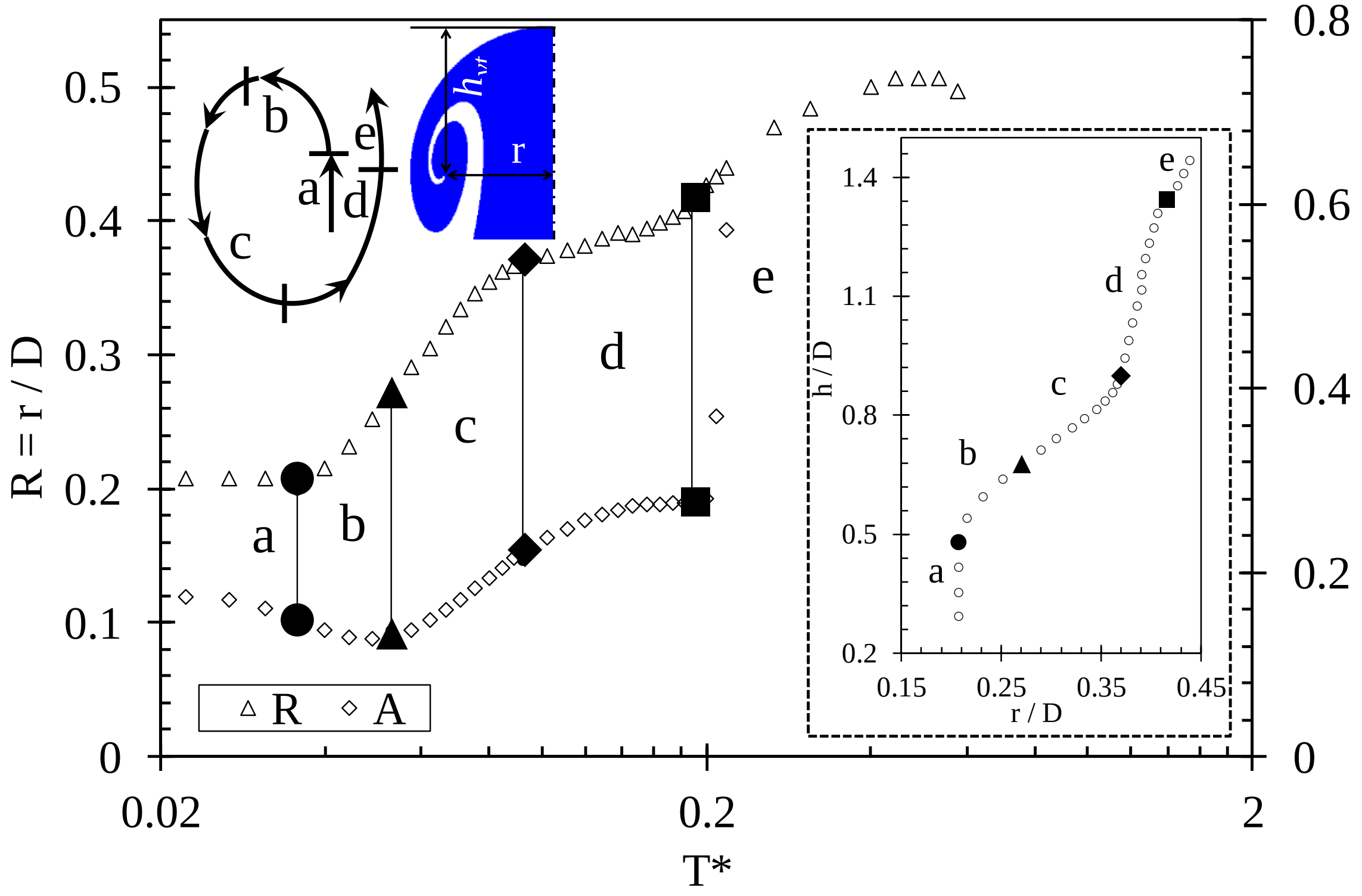

**Figure 6:** The axial and radial growth of the tip of the inner gas vortex ring over time and space during the stages of cavity collapse of a bubble at the free surface; Different growth stages are mentioned.

**Stage b:** At this stage, the radial outward motion of the entrained air tip is started along with continued axial traversal. As a result, the motion of the entraining gas in radial direction rises gradually, which is reflected in the increasing trend of "R" in figure 6. On the other hand, the decreasing trend of "$h_{vt}$" continues, and towards the end, due to outward bend at point “b”, “$h_{vt}$” reaches to the point of inflection. Due to both radial growth and decreasing “$h_{vt}$”, entrainment tip is found to be moving up and going away from the axis in the spatial plane (inset).

**Stage c:** At this stage, the tip of entrained air inside vortex starts moving downward, opposite to the rising inner gas. But, due to continuous entry of air inside the vortex ring through the base, radial growth (R) of the entrainment tip continues along with the enlarging size of the vortex ring. Due to the downward motion of the entrainment tip, “$h_{vt}$” increases, and slow upward motion in an overall axial distance of the tip (in inset) can be observed at this stage.

**Stage d:** In this stage, radial growth is almost constant, and the tip of the vortex ring is steadily heading upward. Although the motion of the entrainment tip is typically inward in this stage, the overall growth of the vortex ring is nullifying the same, maintaining a feeble increase of "R" over time. At the beginning of the stage, “$h_{vt}$” increases following the similar trends of stage c, due to parabolic growth of inner gas tip (figure 4). At later part of this stage, during the upward motion of the entrainment tip, “$h_{vt}$” saturates its growth and follows, more or less, constant gap, signifying equivalent upward velocity of entrainment and the inner gas jet tip. At this advanced level of entrainment growth, a sharp rise in the overall axial distance of entraining air tip can be observed in the inset of Figure 6.

**Stage e:** When the top jet drop reaches in the vicinity of the tip of the inner gas jet, the value of $h_{vt}$ rises and finally shoots up after bypass of the drop. The tip of the inner gas jet increases its height as it has to move along with the droplet (Figure 3 (m)). Radial expansion of the vortex ring at this stage also rises following the trend of stage “d” and reaches a maximum. Around the maxima of R, multiple jet drop bypasses through the tip of inner gas and drag a thick vertical stem following the path of the drop (stage “c” in Figure 4). The vortex ring collapses in the radial direction to feed the first moving stem and induces a reduction in R. The falling trend of R has shown the same after reaching the maxima. It needs to be noted that the inset figure of the spatial plane has been plotted up to T* ~ 0.2.

### 3.3. Effect of Bond number (Bo)

After knowing the dynamics of the inner gas for a bursting bubble, it has been felt appropriate to determine the reach of the aroma of the bubble for different sizes. The role of the Bond number on jet/drop velocity has been explained well by Berny et al. [25], Deike et al. [12] from their numerical work, and the similar has been re-iterated in the theoretical work from Ganan-Calvo [22]. But in the present context, for the very first time, we are stressing upon understanding the role of Bo on the dynamics of inner gas.

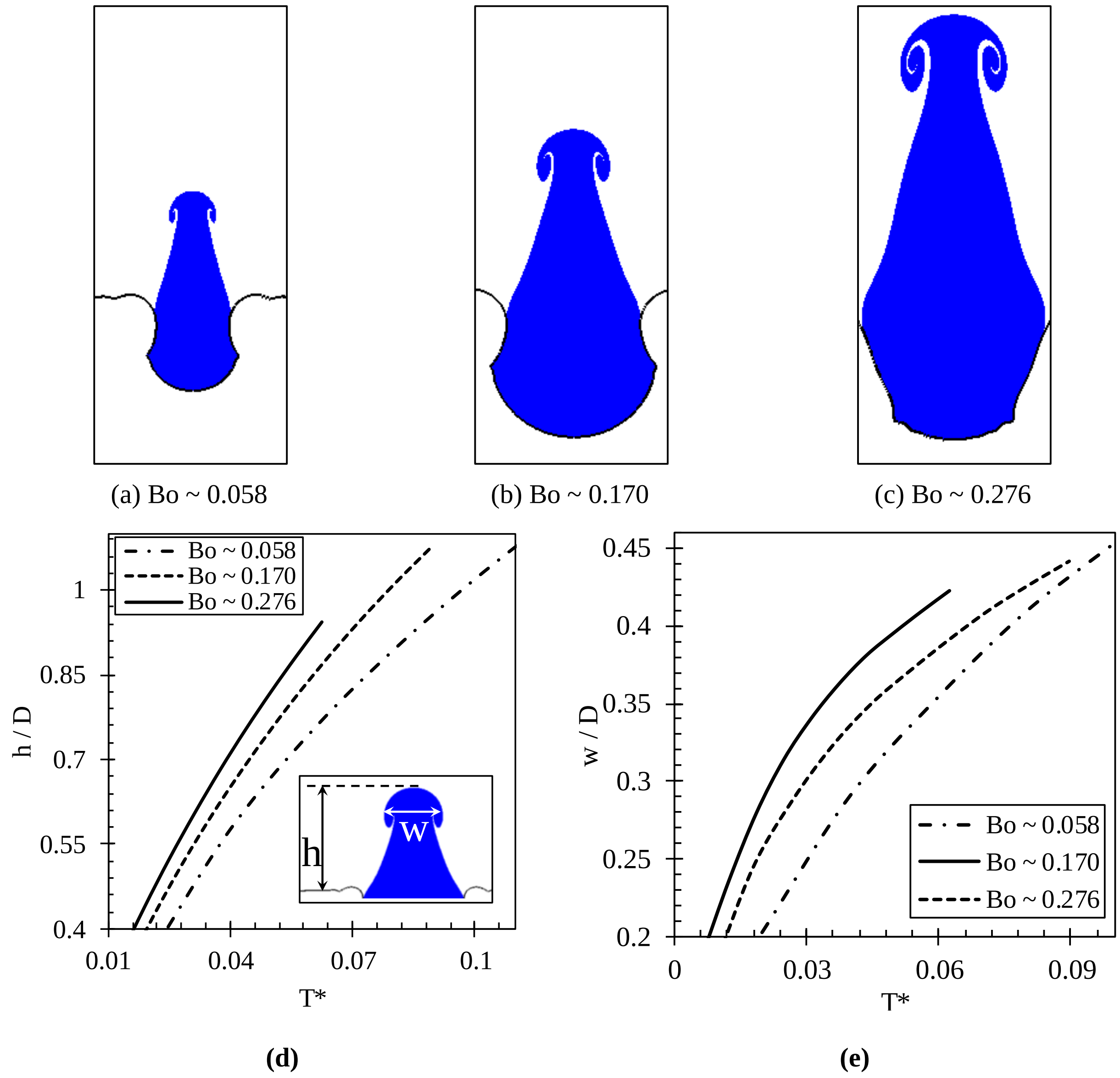

**Figure 7:** (a-c) Phase contours of evacuated inner gas of bubble at T* ~ 0.099 ms for different Bond numbers and (d) & (e) are quantitative comparisons of vertical rise and maximum radial expansion over the unperturbed free surface.

Accordingly, simulations have been performed for a range of Bond number (Bo) with the corresponding static shape of the bubble. Simulated results clearly show that the reach of inner gas in the vertical direction increases with the increasing Bo. This may be due to the rise in pressure within the bubble with increasing Bo, which causes additional pumping force so that the bubble pushes the inner gas up to high heights. A comparative demonstration at some non-dimensional time, T* = 0.099, for Bo as 0.058, 0.170 and 0.276 are shown in Figure 7(a-c). It should also be noted that not only higher heights are reached by inner gas with an increase of Bo, but also the vortex ring size is increased with Bo. Figures 7(d) and 7(e) are displaying the temporal evolution of the height reached by the inner gas jet (h) over the unperturbed free surface and the diametric width of the vortex ring (w) for different Bo values, respectively. It is to be noted for both the illustrations, height, and width have been non-dimensionalized with a characterized length scale of the bubble. The peak height of the inner gas and the stretch of its vertex ring increase with time for the entire range of Bo.

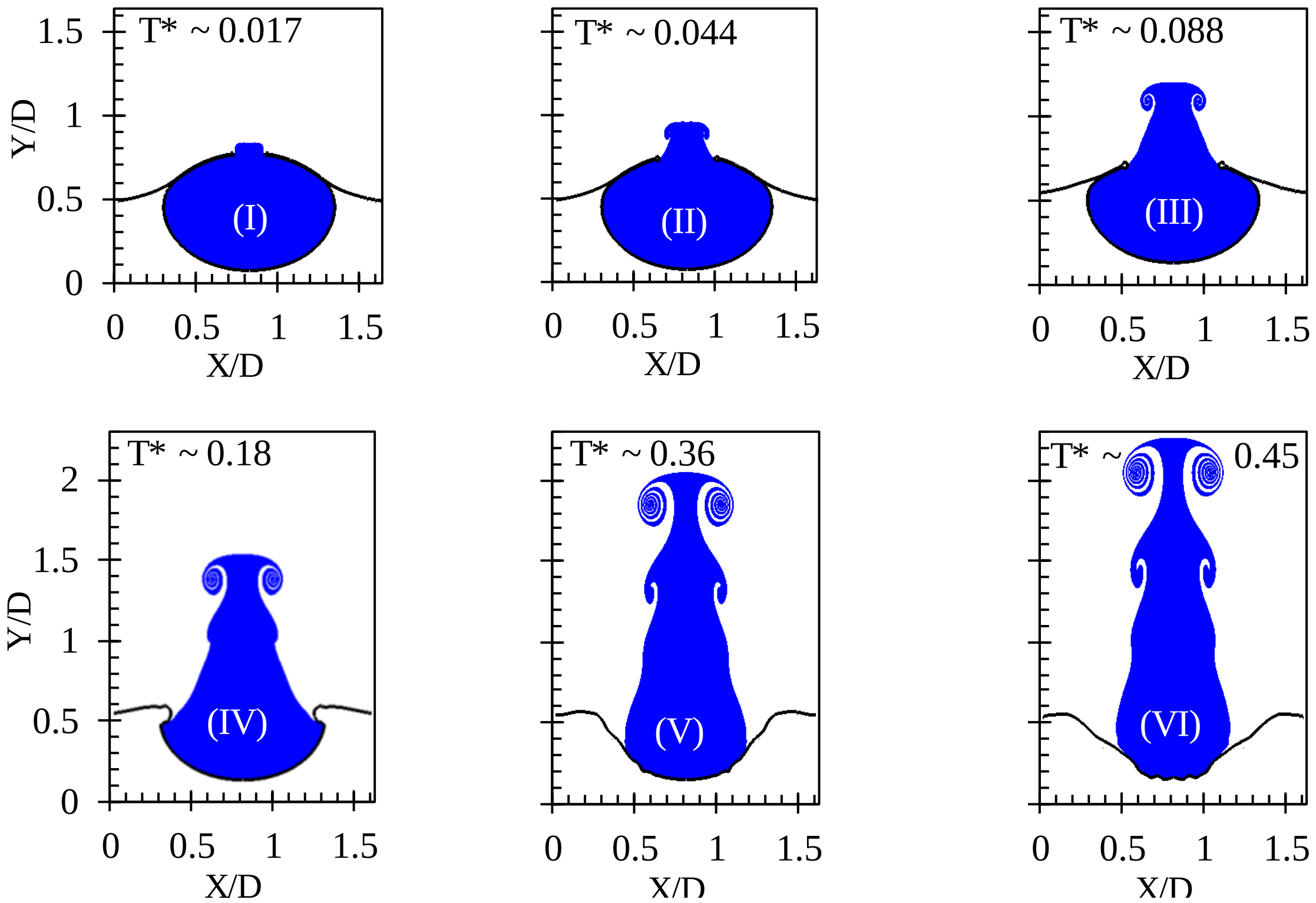

**Figure 8:** Temporal sequences of evacuating inner gas in the case of two distinct vortex rings appeared over the unperturbed free surface for Bo ~ 2.

Air entrainment in upward moving inner gas has been observed in our simulation result of lower Bo as described earlier. The location of entrainment is mainly at the neck of the inner gas jet. The same has been observed approximately until Bo ~ 1. For bubble bursting with static shape and Bo > 1, the tall inner gas jet may entrain at more than one location due to relative velocity between gas streams. For the study, in Figure 8, inner gas dynamics have been reported at Bo ~ 2.

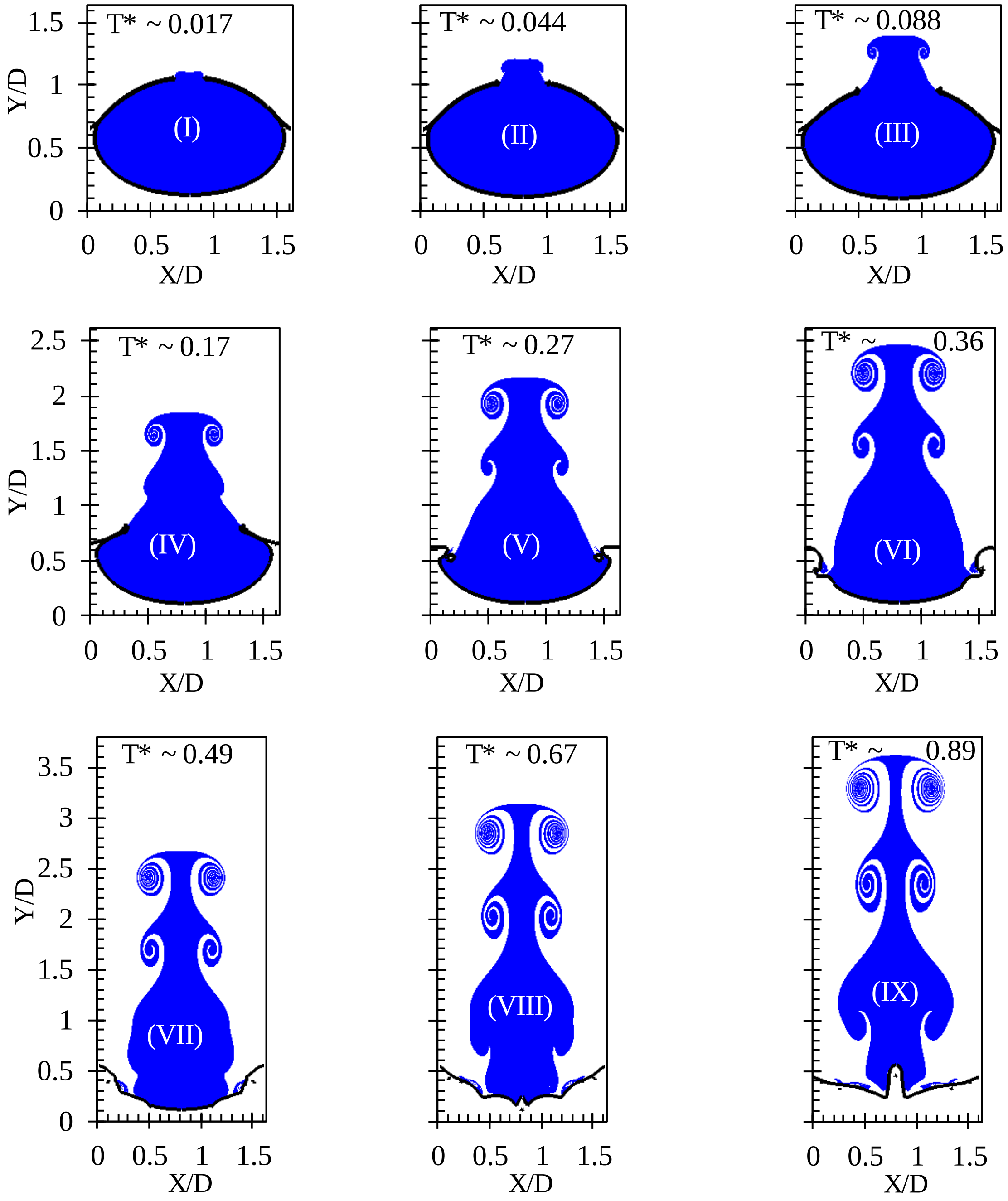

**Figure 9:** Temporal sequences of evacuating inner gas for three distinct vortex rings appeared over the free surface for Bo ~ 4.

One can observe that after continuing till the high non-dimensional time, T* = 0.18, apart from the vortex ring at the neck, another vortex initiates around a heap of the inner gas jet. The circulation of the second vortex is slow at as some time as compared to the first vortex. Along with the first vortex, the second one also moves up and increases air entrainment with time. At Bo ~ 2, a third vortex ring initiates at the knee of inner gas, but its growth is suppressed subsequently. Similar simulation of inner gas dynamics has been done for large bubbles at Bo ~ 4, to take a step further.

One can note in Figure 9 that with an increase of bubble size at Bo ~ 4, the inner gas accommodates the provision of another third entrainment at the knee along with two other locations of air entrainment at the heap and the neck of the inner gas jet. It is evident that entrainment and strength of the topmost vortex are the highest at any time and then slow down respectively in the case of second and third vortex locations. The number of complete circulations of the entrainment in inner gas is highest at the topmost vortex ring and subsequently reducing at lower vortices. Till the lowest vortex ring, it has been observed that entrainment has not even completed one complete circulation. It is to be noted that from these situations of large-sized bubbles (Bo > 1), during bursting, the dwarf liquid jet is being produced, and subsequently, no jet drops are observed, which may have dragged the topmost point of the inner gas jet at higher heights. Similar multi vortices have been reported in experiments by Al Dasouqi and Murphy [30-31], and the observation matches quite will with numerical findings.

### 3.4. Under breathing condition

Next, we have focused on exploring the effect of breathing processes on the bursting bubble aroma. In this section, using the asymmetric numerical setup as described in section 2, we have tried to investigate the dynamics of bubble aroma after bursting when the drinker's respiration airflow influences it. Simulations have been set up for sucking air upward (from the domain top boundary), from the vicinity of the bursting bubble at a rate of 1.4 m/s, which is the approximate velocity of air-breathing in from human beings.

A similar situation has also been studied concerning breathing out, which throws air towards the zone of the bursting bubble at a rate of 1.4 m/s from the top of the domain. The breathing cases are modeled by configuring the inflow and outflow boundary conditions at the top of the domain and above the static shaped bubble. Simulation setup has been decided through a rigorous mesh independence test, as described in Figure 2. The rising inner gas obtained from each case is compared to surrounding quiescent air. Pictorial views of inner gas contour at inhalation, quiescent air, and exhalation situations are shown in Figure 10 (a-c), the comparison has made at T* ~ 0.16. At this time frame, it can be observed that the reach of inner gas in the upward direction is enhanced during inhalation, and inner gas flow is suppressed during exhalation than normal standstill air.

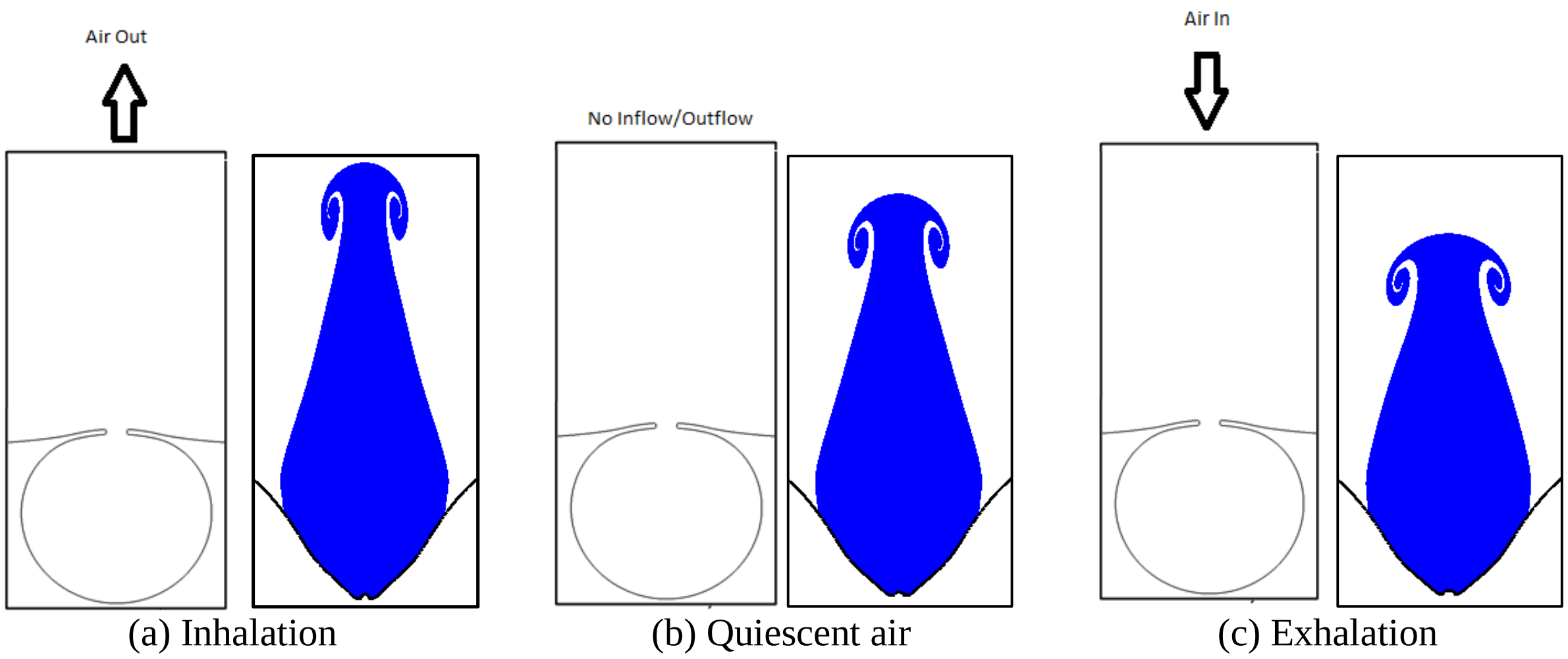

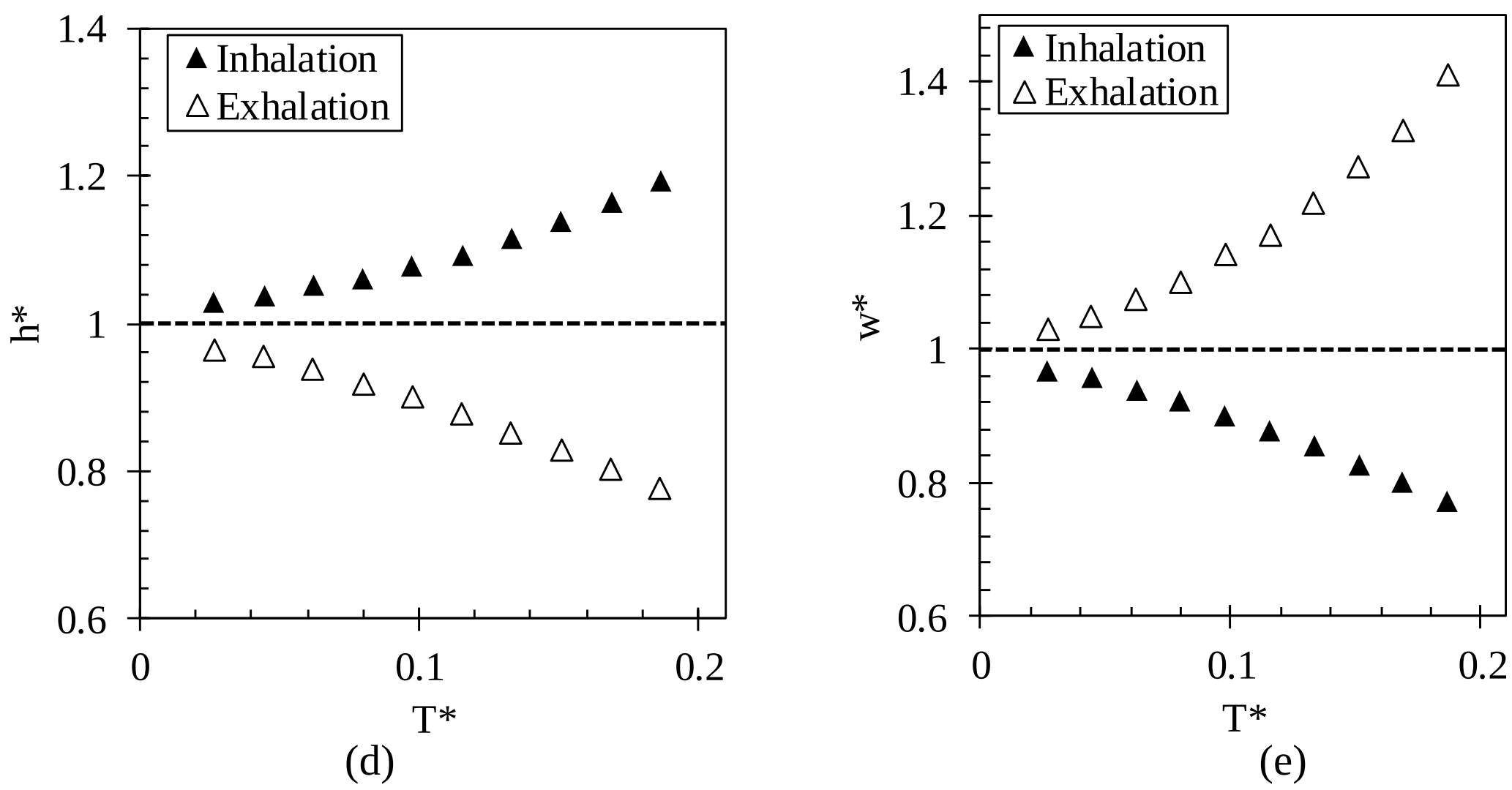

**Figure 10:** (a-c) Phase contours of the inner gas of bursting bubble at T* ~ 0.16; (d) & (e) are the relative axial and radial expansion of inner gas over the unperturbed free surface in inhalation and exhalation with respect to quiescent air situation, respectively

Quantification has also been made from finding out the temporal variation of enhancement and suppression in inner gas reach than still air situation. Figure 10 (d) shows that during inhalation, up to 20% increase in height can be reported till T* ~ 0.2 whereas, on the other hand, more than 23% decrease has been noted in the maximum height of gas jet during exhalation like situations. As time progresses, the effect of breathing in and out will continuously deviate the reach from still air. As the inner gas is being suppressed by airflow during breath out the situation, the restricted inner gas spans in the horizontal direction to make vortex ring diameter larger that still air. The opposite happens in case of breath in the situation. In that case, the vortex becomes weaker than the still air situation. Quantification in Figure 10(e) shows a more than 40% increase in vortex ring diameter due to suppressed inner gas jet than still air situation. On the contrary, the case of breathing in reported around 23% reduction in vortex strength in terms of diameter.

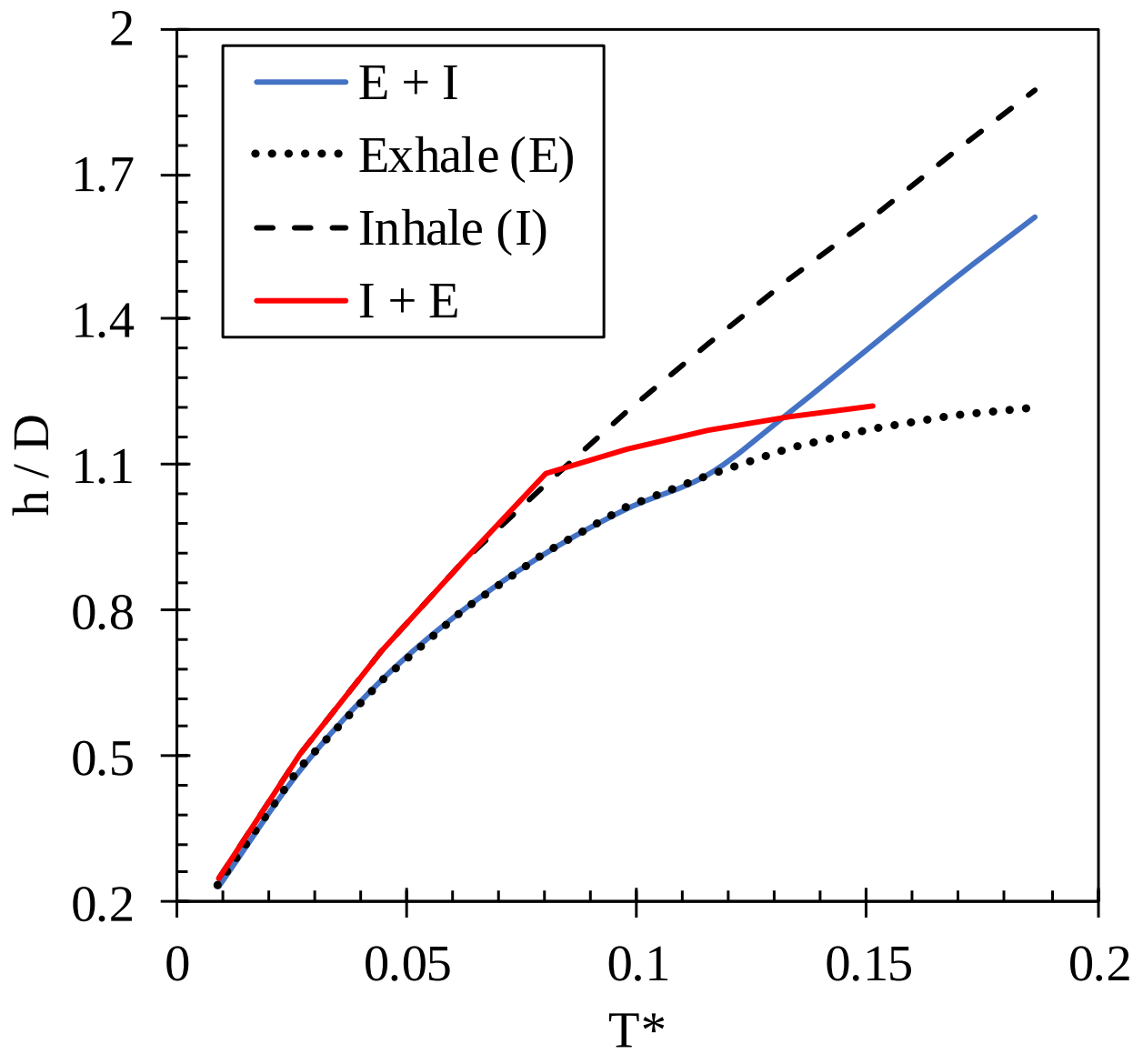

**Figure 11:** Comparison of inner gas of bubble uprise over the unperturbed free surface from separate inhale and exhale with its combination.

Further, extending the same principle of suppressed and enhanced internal gas release as a result of inhalation and exhalation of the human respiration process, an effort has also been made to understand the same during the transition period of inhalation and exhalation. To mimic the same, we have shown two situations; 1) inhalation experiences up to 0.9 ms, and after that, exhalation starts, and 2) exhalation goes up to 1.3 ms, then inhalation begins. The tip of rising inner gas over the unperturbed free surface is obtained and compared with the solitary version of inhalation and exhalation, as shown in Figure 11. In an earlier discussion, it has been stated that the quantity of inner gas gets suppressed in breathing out and enhanced during breathing in situations. Following the same, when inhalation ends, and exhalation begins, rising inner gas experiences sudden fall in the growth rate while reverse in vice versa, as shown in Figure 11. Variations of inner gas jet tip height during pure inhalation and exhalation are also shown in Figure 11 as guiding to the situations of crossover between the inflow and outflow situation of the domain. Change of vertical rise and horizontal spread patterns of inner gas during respiration, over a more extended time, will form an inner gas cloud over the free surface [39].

## 4. Asymmetric static bubble perforation

Once a gas bubble arrives at the free surface, a small hole creates at the bubble cap that initiates the physical phenomenon of the inner gas. From earlier studies, perforation sites are assumed to have solid particles, insoluble proteins, or any other impurities present in the liquid, acting in piercing the bubble cap on the axis of the simulation. In actual scenario, bubble cap piercing does not occur at each instance at the same place, but rather at random. So, to mimic it, we have run three-dimensional simulations in this section by azimuthally setting the specific possible perforation position (θ).

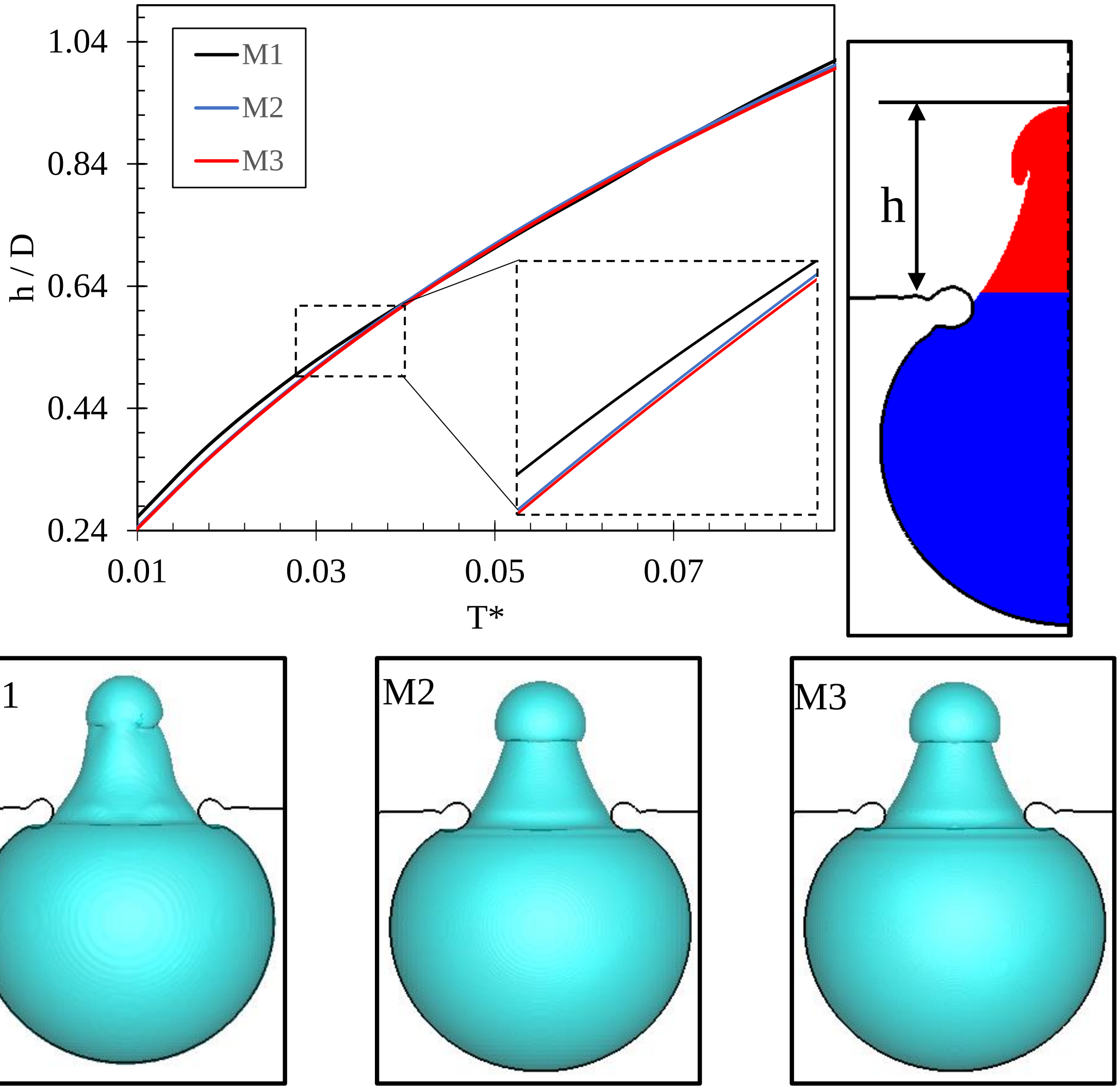

**Figure 12:** Grid independent study performed at several levels of refinement, $T^* = t / \sqrt{\rho_l D^3/\sigma}$

To check the appropriateness of reported results, we conducted three-dimensional simulations at several grid sizes and reported the temporal evolution of uprising inner gas jet tip over the unperturbed free surface during the sequential stages of the bubble bursting in Figure 12. Here, $T^*$ is the nondimensionalized time with a capillary scale. Grid discretization, M2, is considered as a suitable grid size to conduct the simulations after observation saturating performance in a zoomed view of the plot in Figure 12. Furthermore, the quantitative information is tabulated at each category of refinement in Table 2. To establish the grid independence of smaller-scale features in these three-dimensional simulations, we have also plotted the inner gas contours, as obtained from different simulations with different mesh features, reported in Table 2. From the figure, one can ensure the mesh independence of small features in the present 3D simulations. M2 mesh is equivalent to level 13 in Gerris, and dynamic adaptation has been implemented in the simulations considering the gradient of tracer T and $T_i$ as criteria and 13 as a maximum level of adaptive refinement.

**Table 2.** Numerical outcomes by performing the grid-independent study of three-dimensional simulations

| **Level** | **M1** | **M2** | **M3** |
|---|---|---|---|
| **Cell size, µm** | 19.6 | 9.76 | 4.72 |
| **Time, hr** | 4.4 | 35.6 | 506.6 |
| **Average % change in accuracy** | **M1→ M3:** 2.318 | | **M2→ M3:** 0.51 |

The azimuthal location of the hole (θ) is measured from the vertical centroidal axis of the bubble, as shown in Figure 13(d). Schematic representation of asymmetric perforation is given in Figure 13 (a-c), followed in the present study. A comparison of the corresponding inner gas contour of the bursting bubble at the free surface is presented in Figure 13 (c-e). Here, one can see that the vortex ring is no longer axisymmetric, in addition to a bent gas jet. In this section, we are interested in exploring the characteristics of the vortex ring, such as radial (w) and axial (h) growth over time. We carried out a quantitative study and reported features of the vortex in Figure 13. Some parameters are considered to describe the vortex characteristics as follows: $h_i$ and $w_i$ are denoting the relative axial and the radial expansion of the vortex ring in comparison with top punctured bubble ($h_n$, $w_n$), $\forall i \rightarrow l, r$. Until T* ~ 0.19, we found that ~ 25% - 35% fall in relative axial growth when the value of θ went to 18°, as clearly depicted in Figure 14 (a). It goes down linearly and shows no recovery in axial growth. As time progresses, along with axial growth, a vortex ring expands radially. Figure 14 (b) illustrates the relative right-wing radial expansion ($w_r$). It reported a maximum fall in radial expansion of ~ 20% - 30% and showed a marginal recovery (~ 2-3%). Since vortex is not symmetric, so we also performed the same analysis for left-wing. The relative axial development of the left-wing ($h_l$) and the corresponding radial expansion ($w_l$) are plotted in Figure 14 (c-d). These indicate that the values are continuously dominating over θ equal to 0° situation. Relative axial and radial expansion increased to ~ 30% and ~ 50%, respectively until T * ~ 0.19.

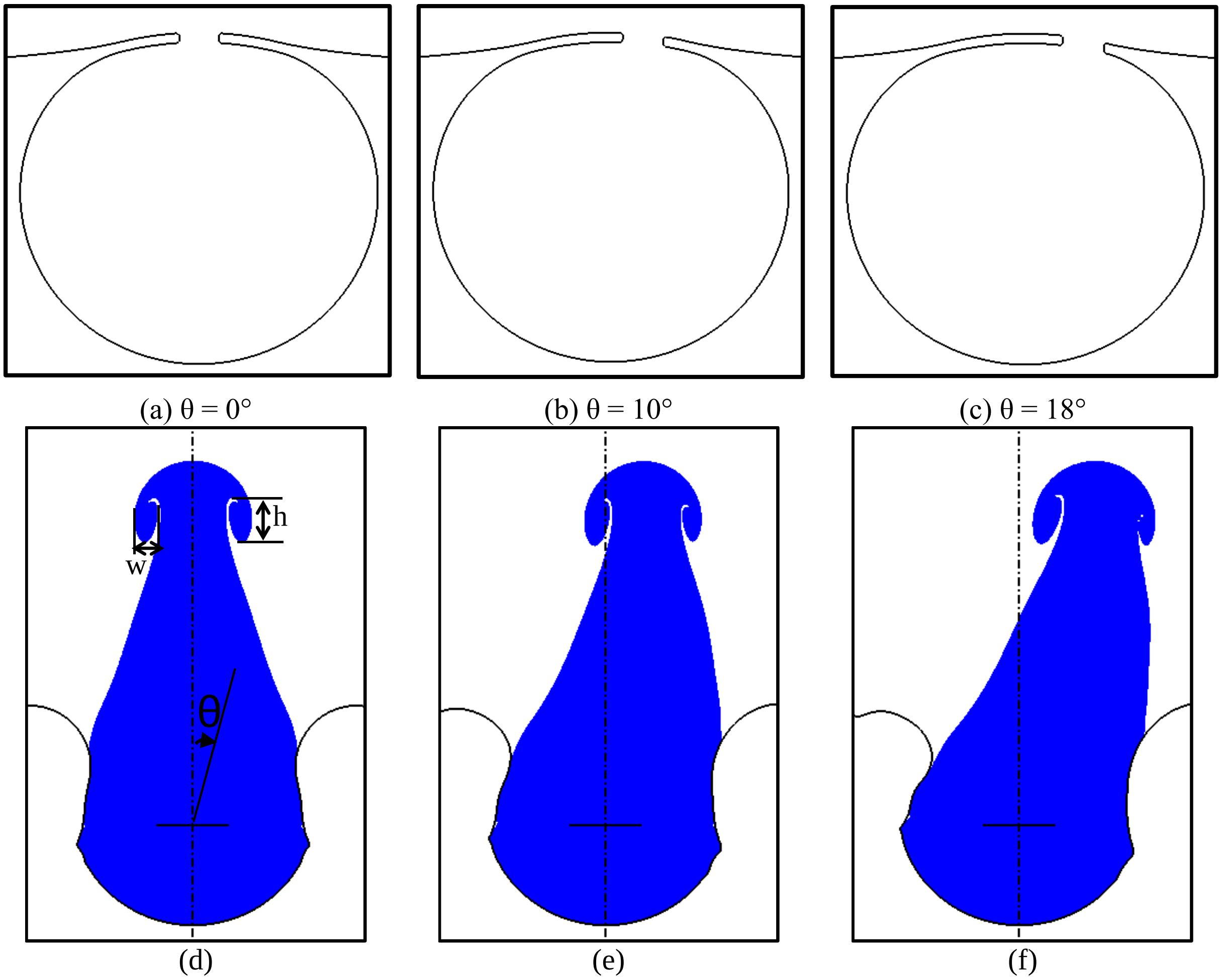

**Figure 13:** (a-c) Schematic representation of asymmetric perforation and corresponding inner gas-phase contours at T* ~ 0.08 of a bubble bursting at the free surface.

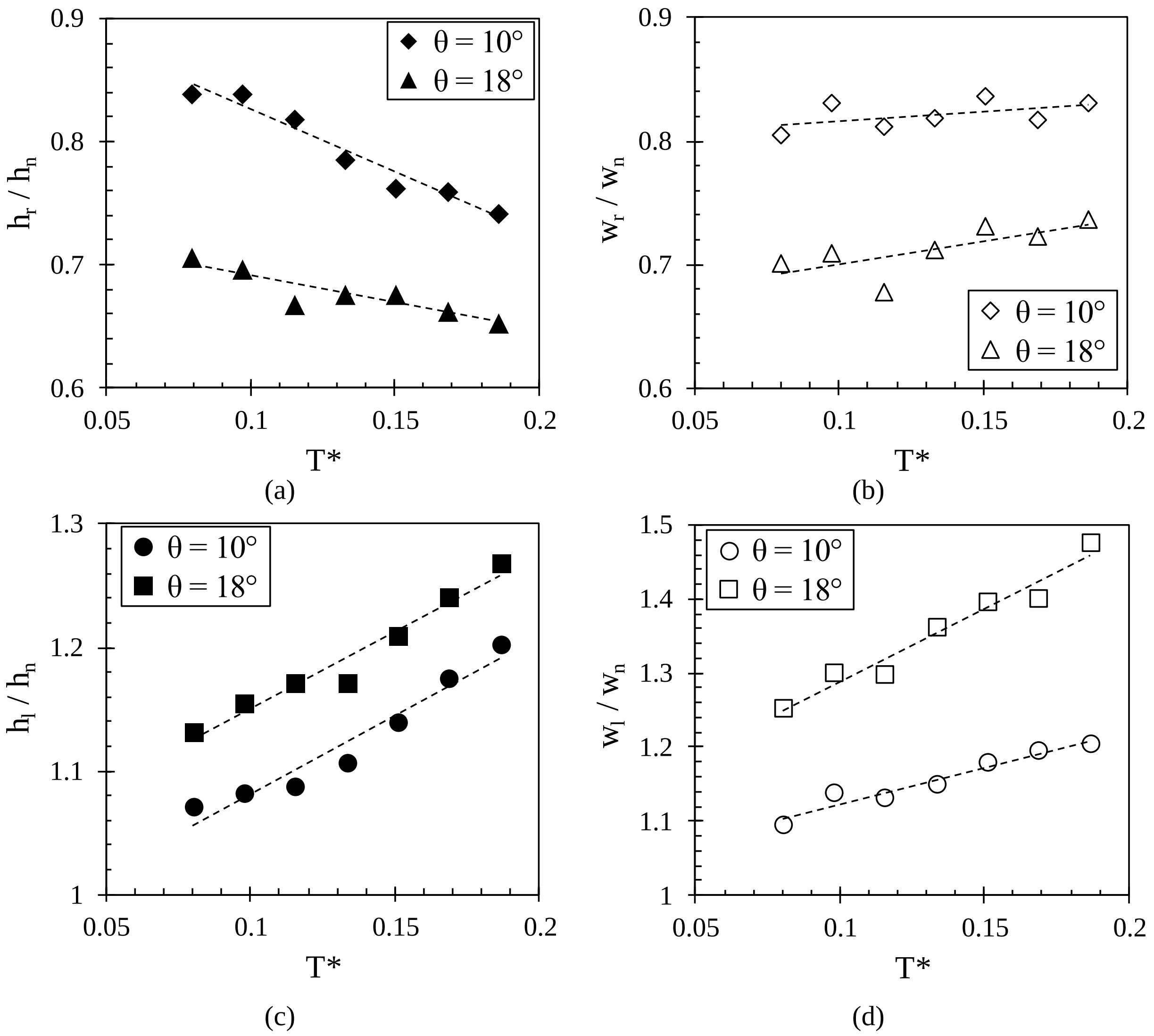

**Figure 14:** Comparison of vortex ring features of uprising inner gas of bubble burst at the free surface, (a-b) axial growth and radial expansion from the wing, and (c-d) corresponding parameters from the left-wing.

Eventually, an effort has also been made to understand the departure of rising inner gas away from the vertical centroidal axis of the bubble under asymmetric bubble cap perforation. Comparative sequences of velocity vectors are provided in figure 15 (a-c) for each asymmetric perforation. These vectors are suggesting that the impulsive flow of the inner gas slowly picks up in space. One can clearly see that a considerable velocity field emerges from the proximity of ripples and went to the maximum at the tip of the gas jet. At θ = 0 °, the velocity field gets inverted at the verge of capillary ripples within the bubble, which indicates the formation of axisymmetric vorticity, and then, moves over the unperturbed free surface parabolically. It is worthy to note that the peak value is lying at the centroidal axis, as shown in Figure 15 (a). As θ increases, the intensity of inverted vectors at the right end capillary decreases, which shows that the strength of vorticity falls relative to left end capillary. The thickness of ripples denoted by the vertical expansion of capillary ripples is also an important parameter to decide the formation of the inclined gas jet during bubble bursting at the free surface. We found that as θ went up, the right end capillary thickness also enhanced, as shown in Figure 15 (a-c).

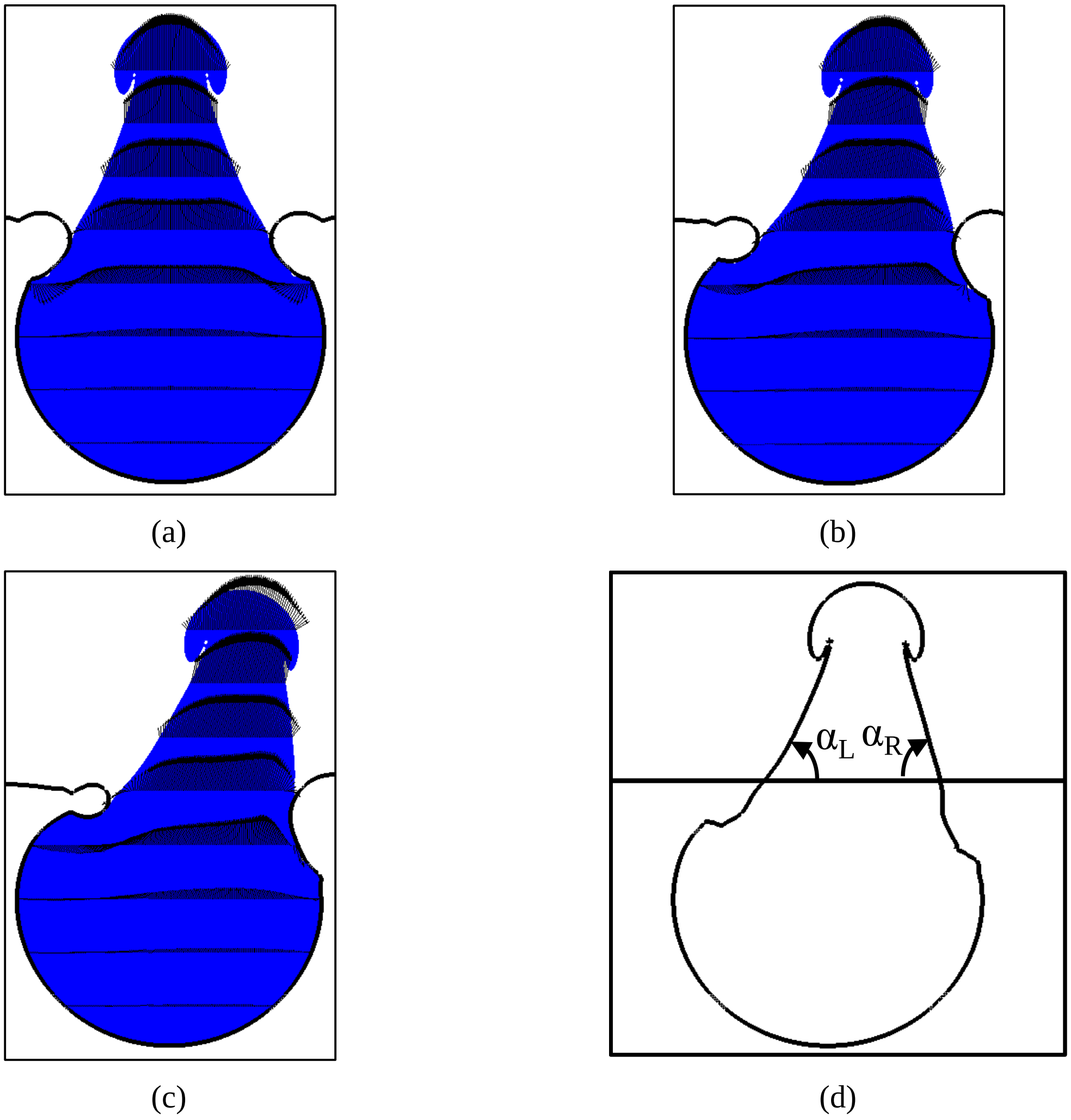

**Figure 15:** Comparison of the velocity vector of uprising inner gas of bubble bursting at the free surface, T* ~ 0.044; the values of θ are: (a) 0◦, (b) 10◦ & (c) 18◦, and (d) Schematic representation of angle subtended by the surface of the inner gas jet with the unperturbed free surface.

During the inner gas evacuation from the bubble cavity, liquid rim acts as a guide to direct the escaping of inner gas through its opening. When an angle, θ = 0°, escaping inner gas is restricted to pass through an axisymmetrically varying orifice; as a result, an axisymmetric gas jet formed. But when $\theta \neq 0°$, the orifice is no longer axisymmetric, which triggers to produce an oblique jet. Figure 15 (d) is shown to depict the angles ($\alpha_L$, $\alpha_R$) subtended by rising gas jet to the unperturbed free surface. Here, the suffix "L" and "R" are used to denote the left and right ends, respectively. These angles ($\alpha_L$, $\alpha_R$) help us in predicting the expansion or contraction of the vortex ring. If $\alpha_L < \alpha_R$, then the gas jet bents towards the right and vortex at the left expands more relative to the right side, and the opposite will happen in vice versa.

## 5. Conclusion:

In the present manuscript, the inner gas of the bursting bubble at the free surface is investigated numerically using Eulerian based volume of fluid approach. The inner gas contours are reported during the sequential stages of the bubble bursting at the free surface. As bubble ruptures at the free surface, the inner gas emerges as a jet from the bubble cavity and subsequently accompanied by the vortex rings. A tiny gas bubble entraps in the liquid pool, after the confluence of ripples inside the bubble cavity. Evacuating inner gas is additionally pushed after the formation of the liquid jet, developing at the base of the bubble cavity. The jet drop, moving fast relative to inner gas jet, drags inner gas and forms a beak-like structure after bypassing its tip.

The axial growth of inner gas over the unperturbed free surface is non-linear until the top jet drop bypasses the gas jet tip. After bypassing, the trend becomes linear, and successive jet drops are also following the same linear trend with lesser strength. Collapsing bubble cavity pumps out the inner gas of the bubble and simultaneously entrains the surrounding air inside the bubble cavity. Quantitative analysis of volumes below the unperturbed free surface reveals that rapid evacuation of inner gas due to high pressure induced by the strong curvature of the bubble in comparison to surrounding gas pressure. Atmospheric air entrainment inside the bubble cavity initially goes at a slower rate and later picks up after free space created during inner gas evacuation. Soon, the whole bubble cavity is found to be occupied only by entrain air, which, finally, is observed to be expelled out during neutralization of the cavity at the free surface. A comparison of the reach of inner gas over the free surface manifests that the rate of expulsion of inner gas reduces upon moving up from the unperturbed free surface.

Inner gas accompanied by the vortex ring entrains the irrotational atmospheric air at the vicinity of the gas jet tip. Bond number (Bo) increases the axial growth of gas jet and radial expansion of vortex rings. The formation of multiple vortex rings is observed upon increasing the Bond number (Bo). In the present study, we found three vortex rings at Bo ~ 4. Under breathing cycles, the axial growth over the free surface gets suppressed and enhanced by exhalation and inhalation process, respectively. In asymmetric perforation of bubble cap, an oblique gas jet emerges out of the bubble cavity, and vortex rings do not hold the planer symmetry, as observed in earlier cases.

**Data Availability**

The data and the numerical codes are available from the corresponding author [Das] for reasonable request.

## References:

1. Andreas, E.L., Edson, J.B., Monahan, E.C., Rouault, M.P. and Smith, S.D., 1995. The spray contribution to net evaporation from the sea: A review of recent progress. Boundary-Layer Meteorology, 72(1-2), pp.3-52.
2. Veron, F., 2015. Ocean spray. Annual Review of Fluid Mechanics, 47, pp.507-538.
3. Blanchard, D.C. and Syzdek, L.D., 1972. Concentration of bacteria in jet drops from bursting bubbles. Journal of Geophysical Research, 77(27), pp.5087-5099.
4. Lewis, E.R. and Schwartz, S.E., 2004. Sea salt aerosol production: mechanisms, methods, measurements, and models (Vol. 152). American Geophysical Union.
5. Kunas, K.T. and Papoutsakis, E.T., 1990. Damage mechanisms of suspended animal cells in agitated bioreactors with and without bubble entrainment. Biotechnology and bioengineering, 36(5), pp.476-483.
6. Boulton-Stone, J.M. and Blake, J.R., 1993. Gas bubbles bursting at a free surface. Journal of Fluid Mechanics, 254, pp.437-466.
7. Al-Rubeai, M., Singh, R.P., Goldman, M.H. and Emery, A.N., 1995. Death mechanisms of animal cells in conditions of intensive agitation. Biotechnology and bioengineering, 45(6), pp.463-472.
8. Liger-Belair, G., 2015. Six secrets of champagne. Physics World, 28(12), p.26.
9. Kientzler, C.F., Arons, A.B., Blanchard, D.C. and Woodcock, A.H., 1954. Photographic investigation of the projection of droplets by bubbles bursting at a water surface. Tellus, 6(1), pp.1-7.
10. MacIntyre, F., 1972. Flow patterns in breaking bubbles. Journal of Geophysical Research, 77(27), pp.5211-5228.
11. Duchemin, L., Popinet, S., Josserand, C. and Zaleski, S., 2002. Jet formation in bubbles bursting at a free surface. Physics of Fluids, 14(9), pp.3000-3008.
12. Deike, L., Ghabache, E., Liger-Belair, G., Das, A.K., Zaleski, S., Popinet, S. and Séon, T., 2018. Dynamics of jets produced by bursting bubbles. Physical Review Fluids, 3(1), p.013603.

13. Cipriano, R.J. and Blanchard, D.C., 1981. Bubble and aerosol spectra produced by a laboratory 'breaking wave'. Journal of Geophysical Research: Oceans, 86(C9), pp.8085-8092.
14. Blanchard, D.C., 1989. The size and height to which jet drops are ejected from bursting bubbles in seawater. Journal of Geophysical Research: Oceans, 94(C8), pp.10999-11002.
15. De Leeuw G., 1986. Size distributions of giant aerosol particles close above sea level. Journal of Aerosol Science, 17, p.293–296.
16. Spiel, D.E., 1997. More on the births of jet drops from bubbles bursting on seawater surfaces. Journal of Geophysical Research: Oceans, 102(C3), pp.5815-5821.
17. Feng, J., Roché, M., Vigolo, D., Arnaudov, L.N., Stoyanov, S.D., Gurkov, T.D., Tsutsumanova, G.G., Stone, H.A, 2014. Nanoemulsions obtained via bubble-bursting at a compound interface. Nature Physics, 10, pp.606–612.
18. Ghabache, E., Antkowiak, A., Josserand, C., Séon, T., 2014. On the physics of fizziness: How bubble bursting controls droplets ejection. Physics of Fluids, 26, pp.121701.
19. Ghabache, E., Liger-Belair, G., Antkowiak, A., Séon, T., 2016. Evaporation of droplets in a Champagne wine aerosol. Scientific Reports, 6, pp.25148.
20. Walls, P.L., Henaux, L., Bird, J.C., 2015. Jet drops from bursting bubbles: How gravity and viscosity couple to inhibit droplet production. Physical Review E, 92, pp.021002.
21. Brasz, C.F., Bartlett, C.T., Walls, P.L.L., Flynn, E.G., Yu, Y.E., Bird, J.C., 2018. Minimum size for the top jet drop from a bursting bubble. Physical Review Fluids, 3, pp.074001.
22. Gañán-Calvo, A.M., 2017. Revision of Bubble Bursting: Universal Scaling Laws of Top Jet Drop Size and Speed. Physical Review Letters., 119, pp.204502.
23. Blanco-Rodriguez, F., Gordillo, J.M., 2020. On the sea spray aerosol originated from bubble bursting jets. Journal of Fluid Mechanics, 886, pp.R2-1-13.
24. Eggers, J., Lai, C.Y., Deike, L. 2018. Bubble bursting: universal cavity and jet profiles. Physical Review Letters, 121, pp.144501.
25. Berny, A., Deike, L., Séon, T. and Popinet, S., 2020. Role of all jet drops in mass transfer from bursting bubbles. Physical Review Fluids, 5(3), p.033605.
26. Liger-Belair, G., Robillard, B., Vignes-Adler, M. and Jeandet, P., 2001. Flower-shaped structures around bubbles collapsing in a bubble monolayer. Comptes Rendus de l'Académie des Sciences-Series IV-Physics, 2(5), pp.775-780.
27. Liger-Belair, G. and Jeandet, P., 2003. Capillary-driven flower-shaped structures around bubbles collapsing in a bubble raft at the surface of a liquid of low viscosity. Langmuir, 19(14), pp.5771-5779.
28. Singh, D. and Das, A.K., 2019. Numerical investigation of the collapse of a static bubble at the free surface in the presence of neighbors. Physical Review Fluids, 4(2), p.023602.

29. Popinet, S., 2003. Gerris: a tree-based adaptive solver for the incompressible Euler equations in complex geometries. Journal of Computational Physics, 190(2), pp.572-600.
30. Al Dasouqi, A. and Murphy, D., 2018. Fluid Mechanics of Gas Jets and Vortex Rings Released from Bursting Bubbles. Bulletin of the American Physical Society, 63.
31. Al Dasouqi, A. and Murphy, D., 2019. Bursting Bubbles and the Formation of Gas Jets and Vortex Rings. Bulletin of the American Physical Society, 64.
32. Fuster, D., Agbaglah, G., Josserand, C., Popinet, S. and Zaleski, S., 2009. Numerical simulation of droplets, bubbles and waves: state of the art. Fluid Dynamics Research, 41(6), p.065001.
33. Tripathi, M.K., Sahu, K.C. and Govindarajan, R., 2015. Dynamics of an initially spherical bubble rising in quiescent liquid. Nature Communications, 6(1), pp.1-9.
34. Kumar, P., Das, A.K. and Mitra, S.K., 2016. Physical understanding of gas-liquid annular flow and its transition to dispersed droplets. Physics of Fluids, 28(7), p.072101.
35. Singh, D. and Das, A.K., 2018. Computational simulation of radially asymmetric hydraulic jumps and jump–jump interactions. Computers & Fluids, 170, pp.1-12.
36. Popinet, S., Gerris Flow Solver, 2013, http://gfs.sourceforge.net/wiki/index.php.
37. Bell, J.B., Colella, P. and Glaz, H.M., 1989. A second-order projection method for the incompressible Navier-Stokes equations. Journal of Computational Physics, 85(2), pp.257-283.
38. Torrey, M.D., Cloutman, L.D., Mjolsness, R.C. and Hirt, C.W., 1985. NASA-VOF2D: a computer program for incompressible flows with free surfaces. NASA STI/Recon Technical Report N, 86.
39. Liger-Belair, G., Bourget, M., Villaume, S., Jeandet, P., Pron, H. and Polidori, G., 2010. On the losses of dissolved CO2 during champagne serving. Journal of agricultural and food chemistry, 58(15), pp.8768-8775.